\documentclass[aps,prb,reprint,superscriptaddress,longbibliography]{revtex4-2}
\usepackage[T1]{fontenc}
\usepackage[utf8]{inputenc}
\usepackage{graphicx}
\usepackage{amsmath,amssymb,bm}
\usepackage{xcolor}
\usepackage{hyperref}
\hypersetup{colorlinks=true,citecolor=blue,linkcolor=blue,urlcolor=blue}

\begin{document}

\title{Tunable Electron Quantum Optics in Graphene Slit Junctions}

\author{Mohamed Amine Rhanbouri}
\email{MohamedAmine.RHANBOURI@um6p.ma}
\affiliation{College of Physical Sciences and Engineering, Mohammed VI Polytechnic University, Ben Guerir, 43150, Morocco}
\author{Wojciech Julian Pasek}
\affiliation{AGH University of Krakow, Faculty of Physics and Applied Computer Science, Department of Applied Informatics and Computational Physics, al. A. Mickiewicza 30, 30-059 Krakow, Poland}
\author{Abdelouahed El Fatimy}
\email{Abdelouahed.ELFATIMY@um6p.ma}
\affiliation{College of Physical Sciences and Engineering, Mohammed VI Polytechnic University, Ben Guerir, 43150, Morocco}

\begin{abstract}
A graphene slit junction is investigated as a compact geometry
for directional electron-beam formation, coherent interference,
and symmetry-controlled spin filtering. Atomistic tight-binding
calculations combined with the nonequilibrium Green's function
formalism show that trigonal warping in monolayer graphene
produces strongly directional electron beams near
$E\simeq2.7\,\mathrm{eV}$. Boundary reflections redirect these
beams toward a common drain, where the transmission exhibits
magnetic-flux-dependent oscillations consistent with an
$h/e$ Aharonov--Bohm period, whose modulation is
characterized by Fourier analysis and harmonic fitting.
A local electrostatic gate provides additional control
of the interference phase.
In separate self-consistent mean-field Hubbard calculations,
the pristine antiferromagnetic ground state is shown to have spin-degenerate transmission, whereas structural or electrostatic breaking of
the left--right symmetry enables spin filtering.
Complementary valley-resolved calculations using a four-band
continuum model demonstrate analogous beam formation in bilayer
graphene at $E=50\,\mathrm{meV}$.
The results identify the slit junction as a common geometry
for electron-optical and spin-dependent transport and motivate
further investigation of low-energy bilayer implementations.
\end{abstract}

\maketitle

\section{Introduction}

High-mobility two-dimensional electron gases (2DEGs) in GaAs/AlGaAs heterostructures have established electron quantum optics as a solid-state counterpart of photonic interferometry, enabling coherent beam splitting, phase control, and interference in mesoscopic conductors \cite{ref1,ref2,ref3}. Graphene provides a complementary platform with several distinctive advantages, including a linear low-energy dispersion, weak intrinsic spin--orbit coupling, and reduced electron--phonon scattering. These properties support long mean free paths and phase-coherence lengths, particularly in high-quality suspended or encapsulated devices \cite{ref4,ref5,ref6}.

These characteristics have enabled a broad range of electron-optics phenomena in graphene, including Klein tunneling and Fabry--P{\'e}rot interference in electrostatically defined $p$--$n$ junctions \cite{ref7,ref8}, as well as Aharonov--Bohm (AB) interferometry in the quantum Hall regime \cite{ref9}. Additional effects such as negative refraction \cite{ref10}, snake states along $p$--$n$ interfaces \cite{ref11}, and anyon braiding in fractional quantum Hall interferometers \cite{ref12} further highlight the richness of coherent transport in this material.

A distinctive opportunity in graphene arises from its valley degree of freedom associated with the inequivalent K and K$^\prime$ points of the 2D Brillouin zone. In monolayer graphene, trigonal warping distorts the isoenergy contours and generates strongly anisotropic group velocities, allowing the formation of highly directional valley-polarized electron beams separated by $60^\circ$ \cite{ref16}. From an experimental perspective, this regime is more readily accessible in bilayer graphene (BLG), where the skew interlayer hopping $\gamma_3$, which couples the
non-dimer sites of the two layers \cite{refMcCann}, produces strongly anisotropic contours at much lower, electrostatically accessible energies
\cite{refMcCann,ref40}. Gate-defined BLG channels have consequently emitted coherent jets whose valley polarization is supported by band-structure and transport modelling \cite{ref17,ref18}, making BLG a
natural setting for valley-dependent beam formation and coherent beam recombination.

Despite these advances, realizing a single graphene geometry that combines several electron-optics functionalities, such as beam splitting, coherent interference, and valley-selective transport, remains an open challenge. An important step in this direction was provided by theoretical proposals based on crossed graphene nanoribbons intersecting at a relative angle of $60^\circ$ \cite{ref13}. Such junctions provide a particularly direct solid-state analogue of an optical beam splitter, enabling the coherent partition of an incoming electronic wave into two outgoing channels. This crossed-junction geometry was later extended to include interaction-induced spin-dependent scattering and spin-polarizing behavior \cite{ref14}, and was further proposed as a building block for AB and Mach--Zehnder interferometers \cite{ref15}. Taken together, these studies established nanoribbon crossings as a powerful platform for graphene electron optics. However, their experimental realization remains particularly challenging, since it requires simultaneous control of ribbon orientation, edge termination, and intersection geometry with near-atomic precision. This motivates the search for alternative compact structures capable of combining similar functionalities within a single device.

In this work, the graphene slit junction is proposed as such an alternative geometry. Rather than relying on the intersection of two separately defined ribbons, the slit selects, within a single flake, the valley-polarized propagation directions set by trigonal warping, while the device boundaries recombine the resulting jets into a two-path interferometer. The zigzag segments created by the slit additionally host edge magnetism, so that the same geometry can act as a spin filter once its left--right symmetry is broken. To the best of our knowledge, this is the first study to propose a graphene slit-junction architecture that combines valley-dependent beam formation, coherent interference, and symmetry-controlled spin filtering within a single device geometry.

The monolayer transport calculations presented here are intended as a mechanism-level study rather than as a quantitative model of a specific BLG device. Although
jetting has been observed in BLG \cite{ref17,ref18}, atomistic simulation of an experimentally sized BLG device remains computationally demanding (Appendix~\ref{app:bilayer}). The beam formation and recombination are governed primarily by the Fermi-contour geometry and the associated group-velocity anisotropy. Monolayer graphene provides a computationally tractable atomistic framework in which these kinematic ingredients and the physical slit edges can be resolved simultaneously. To examine the connection to BLG at low energy, a four-band continuum model is implemented in a device with the same slit layout but larger dimensions. The resulting valley-resolved current maps demonstrate directional beam formation at $E=50\,\mathrm{meV}$, as discussed in Appendix~\ref{app:bilayer}.

The paper is organized as follows. Section~II introduces the tight-binding Hamiltonian and the nonequilibrium Green's function formalism used to compute the transport properties. Section~III
presents valley-dependent beam formation, interference, spin filtering, and the effects of disorder. Section~IV discusses experimental considerations and the relation to BLG, and Sec.~V summarizes the main conclusions. Appendix~\ref{app:bilayer} compares monolayer and bilayer models and presents the low-energy BLG beam-formation calculations.
Appendix~\ref{app:jets} discusses electrode characterization and multi-electrode detection of interference and diffraction.
Additional numerical details and control calculations are provided in the Supplemental Material (SM)~\cite{SM}.\nocite{Delplace2010,Rhim2012,Chakraborti2024,Ge2024,Low2009}

\section{Methods}

\begin{figure}[htbp]
\centering
\includegraphics[width=0.46\textwidth]{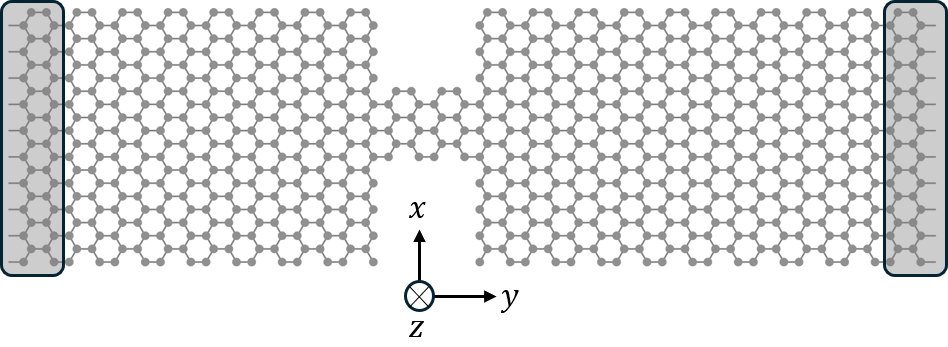}
\caption{Device geometry of the graphene slit junction. An armchair graphene nanoribbon with a narrow central slit is connected to two semi-infinite metallic electrodes on the left and right. The graphene sheet lies in the $xy$ plane, with $z$ normal to the plane; $x$ and $y$ are chosen perpendicular to the armchair and zigzag boundaries, respectively.}
\label{fig:device}
\end{figure}

The device geometry is shown in Fig.~\ref{fig:device}. The graphene layer lies in the $xy$ plane, with the $z$ direction normal to the sheet. Throughout this work, $x$ is chosen as the in-plane direction perpendicular to the armchair boundaries and $y$ as the in-plane direction perpendicular to the zigzag boundaries. The device consists of an armchair graphene nanoribbon with a small slit junction at its center, connected to two semi-infinite electrodes on the left and right. The electrodes are taken sufficiently wide to be considered metallic (see Appendix~\ref{app:jets}). Based on this geometry, the model Hamiltonian of the system is defined and used to compute the transport properties.

\subsection{Model Hamiltonian}

The single-particle tight-binding Hamiltonian for the $\pi$ electrons with a single $p_z$ orbital per site can be written as
\begin{equation}
H = \sum_{\langle i,j\rangle} \left(t_{ij}c_i^{\dagger}c_j + \mathrm{H.c.}\right) + \sum_i \varepsilon_i c_i^{\dagger}c_i,
\label{eq:tb}
\end{equation}
where $c_i^{\dagger}$ ($c_i$) creates (annihilates) an electron on site $i$. In the absence of a magnetic field, the hopping parameter is taken as $t_{ij}=t=-2.7\,\mathrm{eV}$, a standard parameterization
of the graphene $\pi$-band tight-binding model
\cite{CastroNeto2009,ref23}. In the presence of a perpendicular magnetic field, the Peierls substitution method \cite{ref19} is used. The hopping matrix elements then acquire a phase factor, $t_{ij} \rightarrow t_{ij}e^{i\phi_{ij}}$, where the phase shift is given by
\begin{equation}
\phi_{ij} = \frac{e}{\hbar}\int_{\mathbf r_i}^{\mathbf r_j} \mathbf A\cdot d\mathbf r.
\label{eq:peierls}
\end{equation}
Using the Landau gauge \mbox{$\mathbf A = -By\,\hat{\mathbf x}$}, with site vectors \mbox{$\mathbf r_i=(x_i,y_i,z_i)$} leads to
\begin{equation}
\phi_{ij} = \frac{\pi B}{\Phi_0}(x_i-x_j)(y_i+y_j),
\label{eq:phasebond}
\end{equation}
where $\Phi_0=h/e$ is the flux quantum.
A magnetic flux $\Phi=BS$ threading an enclosed area
$S$ produces the AB phase shift
\begin{equation}
\Delta\varphi = 2\pi\frac{BS}{\Phi_0}
= 2\pi\frac{\Phi}{\Phi_0},
\label{eq:abphase}
\end{equation}
in terms of which Eq.~(\ref{eq:phasebond}) becomes
\begin{equation}
\phi_{ij} = \frac{\Delta\varphi}{2S}
(x_i-x_j)(y_i+y_j).
\label{eq:phasebondab}
\end{equation}
The effect of the magnetic field outside the device region is ignored; that is, the Peierls phases are not included in the leads \cite{ref15}.
Zero onsite energy ($\varepsilon_i=0$) in the pristine device has been assumed. To evaluate the robustness of the device against disorder,  the Anderson model is employed, where the onsite energies are sampled from a uniform random distribution of width $W$ \cite{ref20},
\begin{equation}
\varepsilon_i \in \left[-\frac{W}{2},\,\frac{W}{2}\right].
\label{eq:anderson}
\end{equation}

For spin-polarization effects, the mean-field Hubbard (MFH) model \cite{ref14,ref21} is used, with Hamiltonian
\begin{equation}
H = \sum_{\langle i,j\rangle,\sigma}
\left(t_{ij}c_{i\sigma}^{\dagger} c_{j\sigma}
+ \mathrm{H.c.}\right)
+ U \sum_{i,\sigma} \langle n_{i\bar{\sigma}} \rangle n_{i\sigma},
\label{eq:hubbard}
\end{equation}
where $\sigma \in \{\uparrow,\downarrow\}$ and $n_{i\sigma}=c_{i\sigma}^{\dagger}c_{i\sigma}$ is the number operator. The Coulomb interaction is parametrized by the onsite repulsion $U=3\,\mathrm{eV}$, which is in the typical range that yields good agreement with \emph{ab initio} calculations~\cite{ref14}. The same value has recently been benchmarked against density-functional calculations for the localized spin states of armchair graphene-nanoribbon junctions, a closely related class of geometries, where the resulting magnetic excitation energies were found to depend only weakly on $U$ across the range of $1\text{--}3\,\mathrm{eV}$ \cite{ref38}.

To induce spin filtering or phase delays, electrostatically defined
$p$--$n$ junctions are also considered through a nonzero onsite term.
In this case, the tight-binding Hamiltonian becomes
\begin{equation}
\begin{aligned}
H ={}&
\sum_{\langle i,j\rangle,\sigma}
\left(
t_{ij}c_{i\sigma}^{\dagger}c_{j\sigma}
+\mathrm{H.c.}
\right) \\
&+ U\sum_{i,\sigma}
\langle n_{i\bar{\sigma}}\rangle n_{i\sigma}
+\sum_{i,\sigma}V_i c_{i\sigma}^{\dagger}c_{i\sigma},
\end{aligned}
\label{eq:tbonsite}
\end{equation}
where $V_i$ is the onsite energy at site $i$. Two regions in the
device are considered, labeled L and R, with onsite energies
$V_L$ and $V_R$, respectively, such that
\begin{equation}
V_i =
\begin{cases}
V_L, & i \in L,\\
V_R, & i \in R.
\end{cases}
\label{eq:pn}
\end{equation}
This enables us to model local electrostatic gating and, in particular, the formation of $p$--$n$ junctions.

\subsection{Transport calculation}

Transport properties are calculated using the nonequilibrium Green's function (NEGF) formalism \cite{ref22,refDatta1995}. The ballistic transmission probability between leads $\alpha$ and $\beta$ is given by the Landauer--B\"uttiker formula
\begin{equation}
T_{\alpha\beta} = \mathrm{Tr}\!\left[ \Gamma_{\alpha}
\mathcal{G} \Gamma_{\beta} \mathcal{G}^{\dagger} \right],
\label{eq:landauer}
\end{equation}
where $\mathcal{G}$ is the retarded device Green's function, $\Gamma_{\alpha}=i\left(\Sigma_{\alpha}-\Sigma_{\alpha}^{\dagger}\right)$ is the coupling matrix between the lead and the device, and $\Sigma_{\alpha}$ is the retarded self-energy of lead $\alpha$.

In order to directly visualize electron-beam propagation paths, one approach is to introduce a negatively charged SPM tip that scans the device surface; the resulting tip-position-dependent conductance can then be used to map the electron flow \cite{ref16}. Here, a simpler, yet effective, method is used based on the calculation of bond and vector transmission \cite{ref23}, as follows.

The bond transmission between atoms $I$ and $J$ is defined as the
sum over all orbital transmissions between the orbitals belonging
to these two atoms
\begin{equation}
T_{IJ}(E)=\sum_{i\in I}\sum_{j\in J} T_{ij}(E).
\label{eq:bondtrans}
\end{equation}
In the present single-orbital model, each atom carries one
$p_z$ orbital, so the orbital and atomic indices coincide
($i=I$ and $j=J$), and this sum reduces to a single term.

The corresponding vector transmission at atom $I$ is then obtained
by summing the bond transmissions weighted by the normalized bond
vectors connecting atom $I$ to its neighboring atoms
\begin{equation}
\mathbf{T}_{I}(E)=\sum_{J}
\hat{\mathbf{r}}_{IJ}\,T_{IJ}(E),
\label{eq:vectortrans}
\end{equation}
where $T_{IJ}(E)$ is the bond transmission of
Eq.~(\ref{eq:bondtrans}),
$\mathbf{r}_{IJ}=\mathbf{r}_{J}-\mathbf{r}_{I}$, and
$\hat{\mathbf{r}}_{IJ}=\mathbf{r}_{IJ}/|\mathbf{r}_{IJ}|$,
with $\mathbf{r}_{I}$ and $\mathbf{r}_{J}$ denoting the
atomic coordinates.

Computationally, transmission probabilities from the Green's function using the open-source software TBtrans (v4.1.5) \cite{ref24} are calculated. The data are post-processed and analyzed using the \texttt{sisl} package \cite{ref25}. The self-consistent solutions of the mean-field Hubbard model, Eq.~(\ref{eq:hubbard}), are obtained with the open-source \texttt{hubbard} package \cite{ref21}.

Multi-electrode calculations \cite{ref24} are also performed and the current flowing between electrodes computed using the Landauer--B\"uttiker formalism
\begin{equation}
I = \frac{G_0}{e} \int_{-\infty}^{+\infty}
\left[f_L(E,kT,\mu_L)-f_R(E,kT,\mu_R)\right] T(E)\, dE,
\label{eq:current}
\end{equation}
where $G_0 = 2e^2/h$ is the conductance quantum, and
\begin{equation}
f_{\alpha}(E,kT,\mu_{\alpha})=
\left[1+\exp\!\left(\dfrac{E-\mu_{\alpha}}{kT}\right)\right]^{-1}
\label{eq:fermi}
\end{equation}
is the Fermi--Dirac distribution function of electrode $\alpha$, with chemical potential $\mu_{\alpha}$, temperature $T$, and $k$ the Boltzmann constant.

\section{Results}

\begin{figure*}[!tp]
\centering
\includegraphics[width=\textwidth]{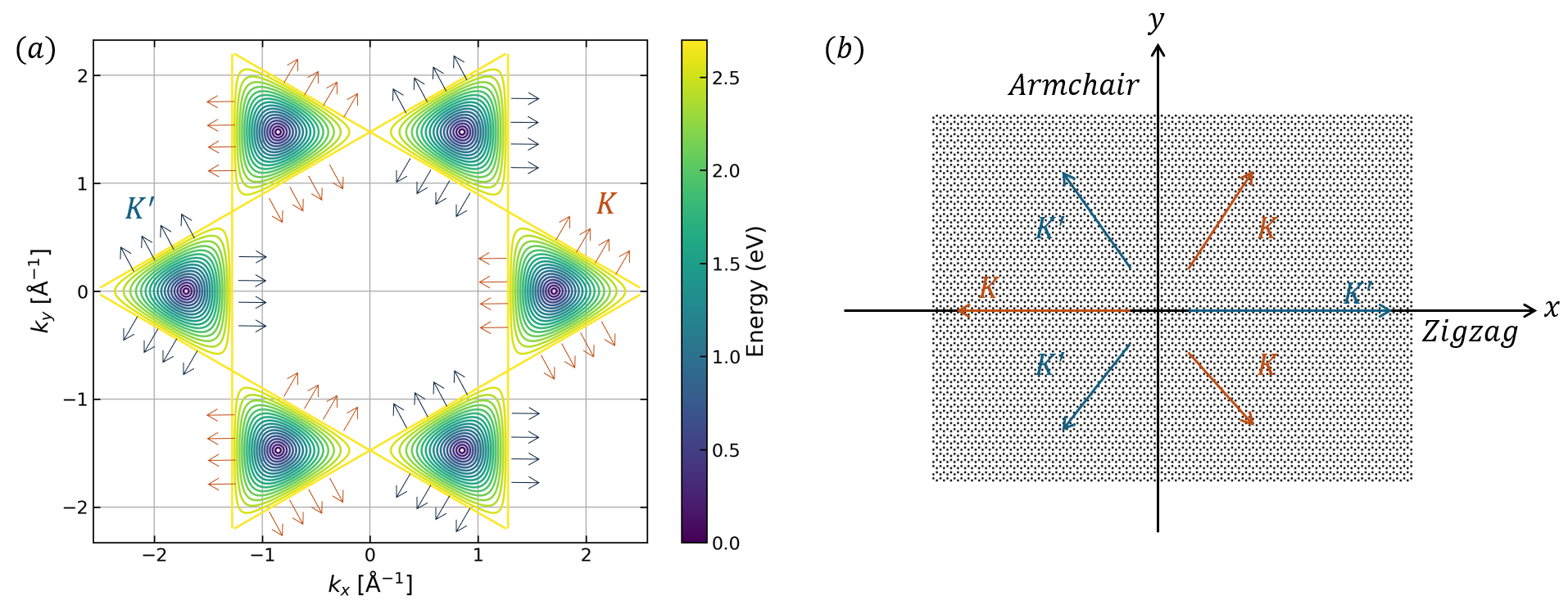}
\caption{Trigonal warping in monolayer graphene. (a) Isoenergy contours from $E=0$ to $E=2.7\,\mathrm{eV}$ in the first Brillouin zone. The group velocities of the K and K$^\prime$ valleys are represented by orange and blue arrows, respectively. (b) Anisotropic group-velocity directions in monolayer graphene at the K and K$^\prime$ points.}
\label{fig:warping}
\end{figure*}

\begin{figure*}[!tp]
\centering
\includegraphics[width=0.9\textwidth]{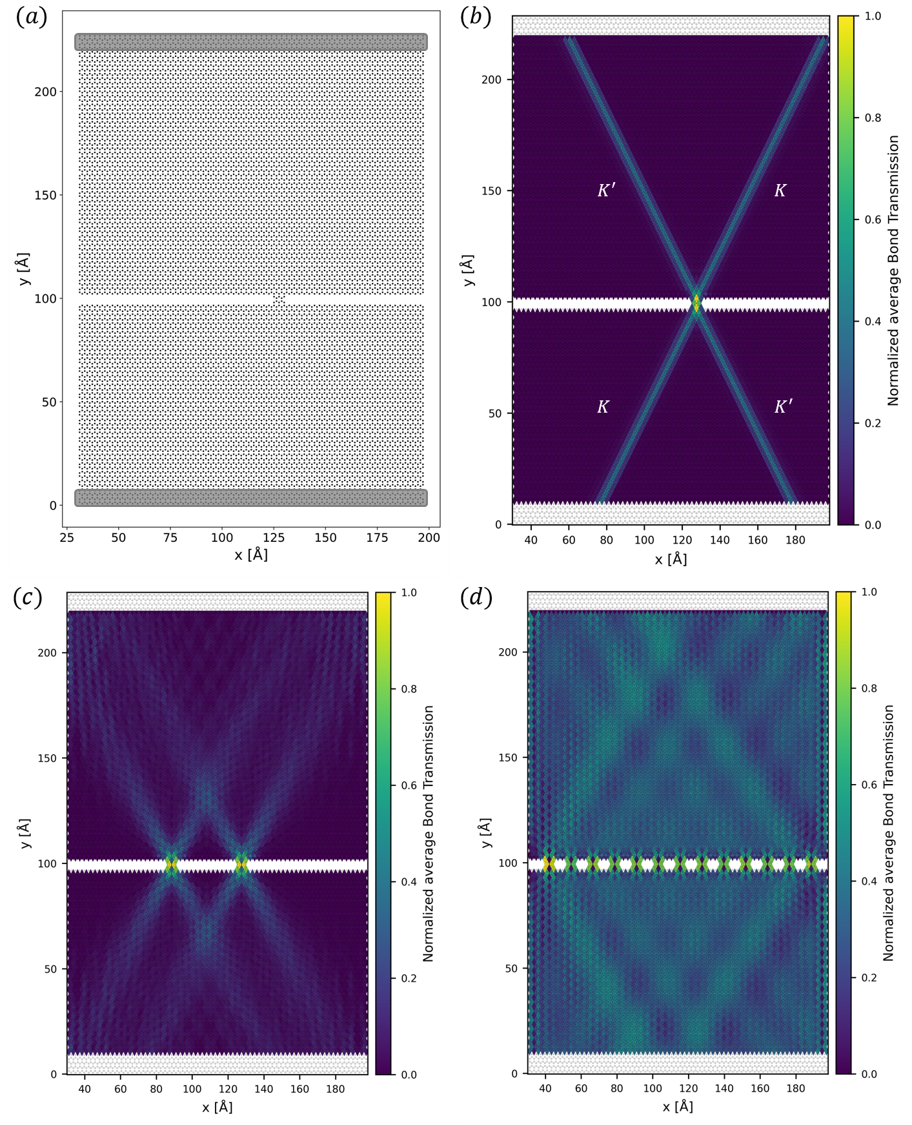}
\caption{Valley-polarized jets, interference, and diffraction in graphene slit junctions. (a) Geometry of the simulated device consisting of an armchair graphene nanoribbon with a small slit in the center; the electrodes are shown in grey. (b) Bond-transmission map at $E=2.7\,\mathrm{eV}$, where trigonal warping is maximal. (c) Bond-transmission map at $E=2\,\mathrm{eV}$ for a double-slit geometry. (d) Bond-transmission map at $E=1\,\mathrm{eV}$ for a multiple-slit geometry, yielding a diffraction-grating-like pattern.}
\label{fig:jets}
\end{figure*}

\begin{figure*}[!tp]
\centering
\includegraphics[width=.95\textwidth]{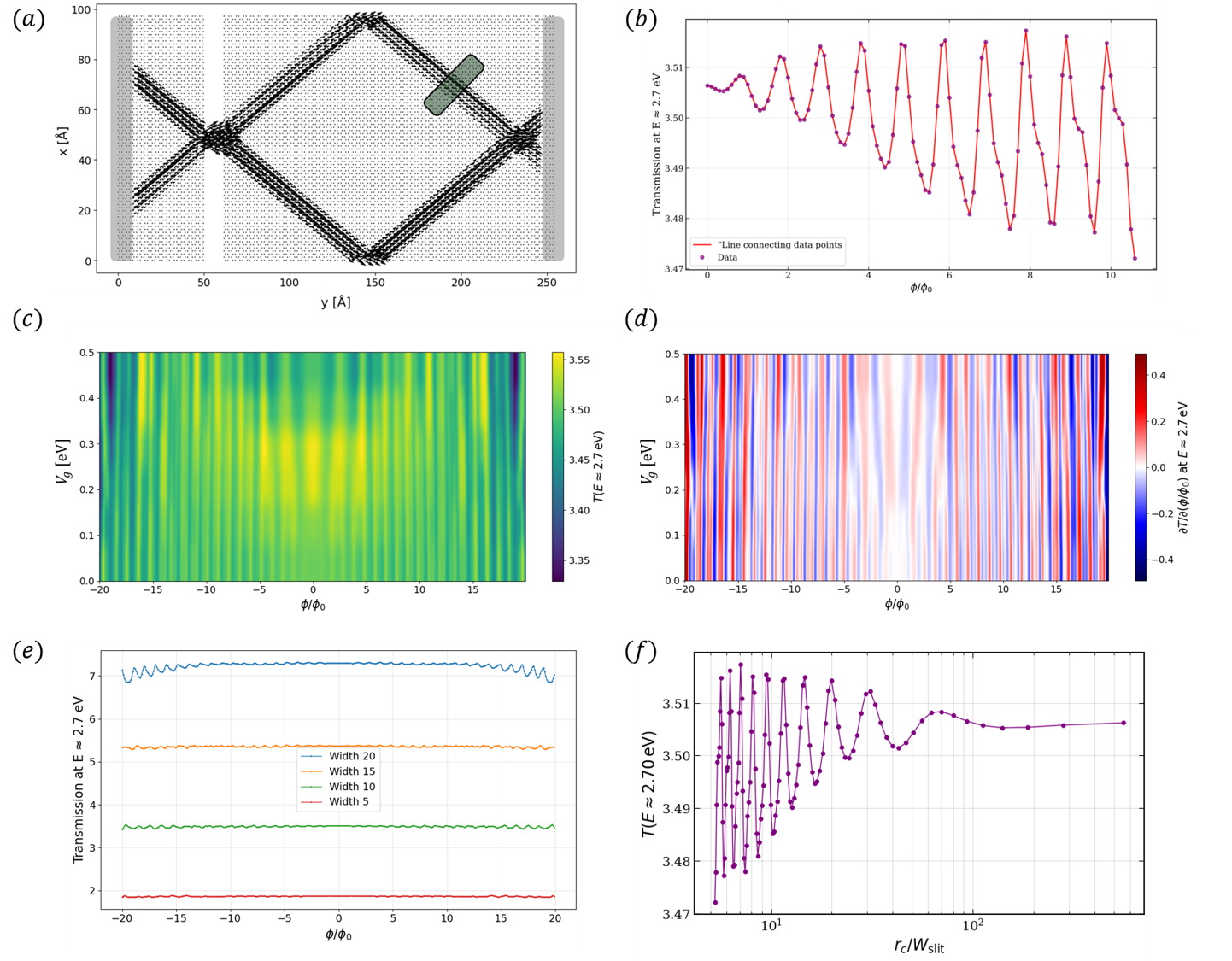}
\caption{Flux- and gate-controlled Aharonov--Bohm interference between valley-polarized electron jets. (a) Vector-transmission map at $E=2.7\,\mathrm{eV}$ for a slit width $W_{\mathrm{slit}}=10\,\text{\AA}$. The two jets leave the slit at an angle of $60^\circ$, reflect from opposite device boundaries, and recombine at the drain. The grey regions denote the electrodes, and the green region on the upper arm denotes the area to which the phase-shifting onsite potential $V_g$ is applied. (b) Transmission at $E=2.7\,\mathrm{eV}$ as a function of normalized magnetic flux $\Phi/\Phi_0$. (c) Transmission at $E=2.7\,\mathrm{eV}$ as a joint function of $\Phi/\Phi_0$ and $V_g$. (d) Flux derivative $\partial T/\partial(\Phi/\Phi_0)$ of the map in (c), which highlights the interference fringes. (e) Transmission versus $\Phi/\Phi_0$ for $W_{\mathrm{slit}}=5$, 10, 15, and $20\,\text{\AA}$. (f) Transmission at $E=2.7\,\mathrm{eV}$ plotted against the dimensionless ratio $r_c/W_{\mathrm{slit}}$, for $W_{\mathrm{slit}}=10\,\text{\AA}$, where $r_c$ is the cyclotron radius at $E=2.7\,\mathrm{eV}$.}
\label{fig:ab}
\end{figure*}

\subsection{Valley-polarized electron jets}

Directional jetting driven by trigonal warping has been observed experimentally in BLG \cite{ref17,ref18}. Atomistic simulation of an entire experimentally sized BLG device remains computationally demanding, and published calculations have therefore used scaled honeycomb lattices \cite{ref39}, four-band continuum Hamiltonians on auxiliary square grids \cite{refChen,ref37}, or atomistic injectors combined with semiclassical propagation \cite{refInglaTEF}. The detailed model hierarchy is reviewed in Appendix~\ref{app:bilayer}. Here, the mechanism is therefore examined atomistically in monolayer graphene,
as motivated in Sec.~I.

In monolayer graphene, this behavior follows from the nearest-neighbor tight-binding dispersion relation \cite{ref16}
\begin{multline}
E = \pm t \left[1 + 4\cos\left( \frac{\sqrt{3}ak_{x}}{2} \right)\cos\left( \frac{3ak_{y}}{2} \right) \right. \\
+ \left. 4\cos^{2}\left( \frac{\sqrt{3}ak_{x}}{2} \right)\right]^{\frac{1}{2}},
\label{eq:dispersion}
\end{multline}
where $a=1.42\,\text{\AA}$ is the carbon--carbon distance.
As shown in Fig.~\ref{fig:warping}(a), the isoenergy contours are circular at low energies, indicating an isotropic group velocity. At higher energies, however, these contours deform due to trigonal warping, reaching maximal anisotropy around $E=2.7\,\mathrm{eV}$. The corresponding group velocities then exhibit valley-polarized directions, as shown in Fig.~\ref{fig:warping}(b). Analytically, these directions can be obtained from the group velocity
\begin{equation}
\mathbf{v}=\frac{1}{\hbar}\nabla_{\mathbf{k}}E\Big|_{E=|t|} =
\begin{cases}
\displaystyle \pm\frac{\sqrt{3}a|t|}{\hbar}\cos\!\left(\frac{3}{2}ak_y\right)(1,0),\\[2ex]
\displaystyle -\frac{\sqrt{3}a|t|}{\hbar}\sin\!\left(\sqrt{3}ak_x\right)\left(\frac{1}{2},\pm\frac{\sqrt{3}}{2}\right).
\end{cases}
\label{eq:velocity}
\end{equation}
Depending on the transport direction, these anisotropies can be selected using slit junctions. In a zigzag nanoribbon, the incident beam splits into three preferred directions, whereas in the armchair case two valley-polarized electron jets emerge. To investigate this in the monolayer case, the geometry shown in Fig.~\ref{fig:jets}(a) is used, consisting of two extended armchair graphene nanoribbons (AGNRs), shown in grey, which serve as electrodes. The ribbons are chosen sufficiently wide to exhibit metallic behavior [see Appendix~\ref{app:jets}, Fig.~\ref{fig:jets_device}(a)]. The device region consists of an AGNR with a narrow central slit, whose width is taken small in order to reduce the number of injected channels \cite{ref16}.

Since the transport direction is along the armchair axis, two valley-polarized electron jets are expected, as predicted in Fig.~\ref{fig:warping}(b). The bond-transmission map at $E=2.7\,\mathrm{eV}$ in Fig.~\ref{fig:jets}(b) confirms this expectation: two well-defined electron beams emerge, separated by an angle of $60^\circ$. In the zigzag case, Fig.~\ref{fig:warping}(b) predicts three outgoing beams. This behavior is also verified using the bond-transmission map shown in Appendix~\ref{app:jets} [Fig.~\ref{fig:jets_device}(b)], in agreement with the results reported in Ref.~\cite{ref16}.

This effect can be further exploited to generate interference between two valley-polarized jets. Figure~\ref{fig:jets}(c) shows the bond-transmission map for a double-slit geometry at $E=2\,\mathrm{eV}$, where trigonal warping is weaker. A clear interference region is observed, from which additional insight into the underlying behavior can be extracted (see Appendix~\ref{app:jets}, Figs.~\ref{fig:double_slit_detection} and \ref{fig:multielectrode_detection}). Finally, Fig.~\ref{fig:jets}(d) shows that a multiple-slit geometry produces a diffraction-grating-like pattern in the bond-transmission map.

\subsection{Aharonov--Bohm interferometer}

Early AB oscillations in graphene were measured in etched and side-gated rings \cite{ref27,ref28}, and the influence of Lorentz-force deflection in such devices was subsequently analyzed theoretically \cite{ref26}. In BLG, gate-defined cavities have exhibited ballistic Fabry--P{\'e}rot interference with a gate-tunable Berry phase and anti-Klein transmission \cite{refLiuBLGFP,refLiuAntiKlein}. A fully gate-defined BLG ring subsequently displayed a dominant $h/e$ period together with the higher harmonics $h/2e$, $h/3e$, and $h/4e$ \cite{refIwakiriBLGRing}, while a continuum-derived four-band square-grid model reproduced the $h/e$ response of a realistic gate-defined BLG quantum ring \cite{refChen}. More recently, graphene quantum Hall interferometers based on $p$--$n$ junctions or gate-defined cavities have reached high visibility, although they require strong perpendicular magnetic fields \cite{ref9,ref29,ref30}.
Graphene Fabry--P{\'e}rot interferometers have also enabled measurements of anyon braiding in the fractional quantum Hall regime \cite{ref12}. A different theoretical architecture uses four zigzag graphene nanoribbons arranged in two parallel pairs and intersecting at $60^\circ$, discussed in Sec.~I, to produce Mach--Zehnder-like interference \cite{ref13,ref14,ref15}.

Here, an alternative device is proposed in which the slit separates the valley-polarized electron jets and the boundaries redirect them toward the drain. A perpendicular magnetic field changes the relative phase accumulated along the two paths, producing an interference pattern in the transmission. The vector-transmission map in Fig.~\ref{fig:ab}(a), evaluated at $E=2.7\,\mathrm{eV}$, shows the two jets leaving the slit at an angle of $60^\circ$, reflecting from opposite boundaries, and recombining at the drain electrode. The corresponding transmission in Fig.~\ref{fig:ab}(b) oscillates with the normalized magnetic flux $\Phi/\Phi_0$.

\begin{figure*}[!tp]
\centering
\includegraphics[width=0.8\textwidth]{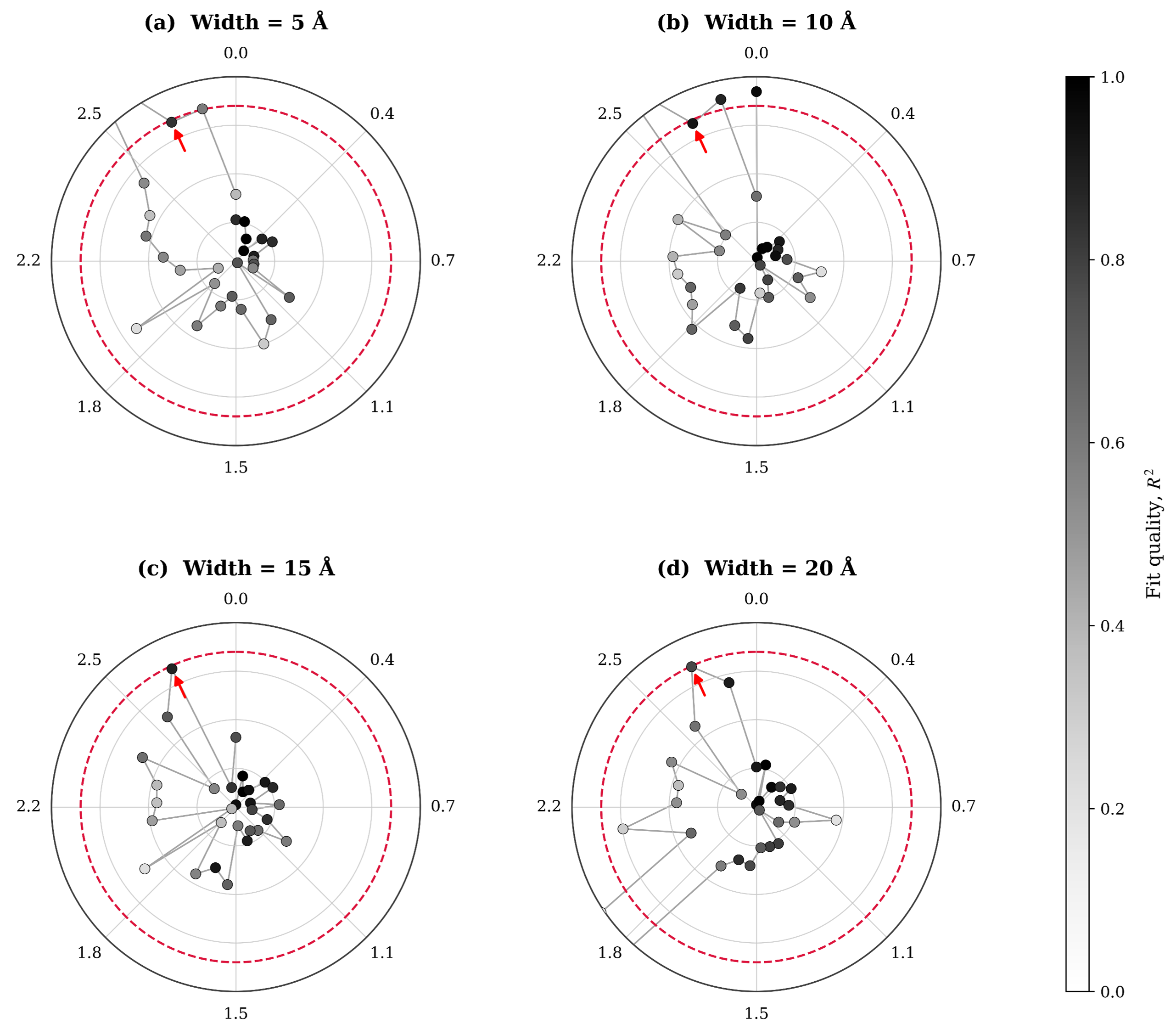}
\caption{Extracted AB oscillation frequency and fit quality as functions of energy given in eV. The angular coordinate represents the electron energy in eV, and the radial coordinate represents the fitted normalized frequency $f$. The grayscale encodes $R^2$, ranging from white ($R^2=0$) to black ($R^2=1$). The red dashed circle marks the $h/e$ expectation $f=1$, while short red arrows identify the sampled points nearest $E=2.7\,\mathrm{eV}$. Results are shown for slit widths (a) $W_{\mathrm{slit}}=5\,\text{\AA}$, (b) $10\,\text{\AA}$, (c) $15\,\text{\AA}$, and (d) $20\,\text{\AA}$.}
\label{fig:ab_freq_energy}
\end{figure*}

\begin{figure*}[!tp]
\centering
\includegraphics[width=.9\textwidth]{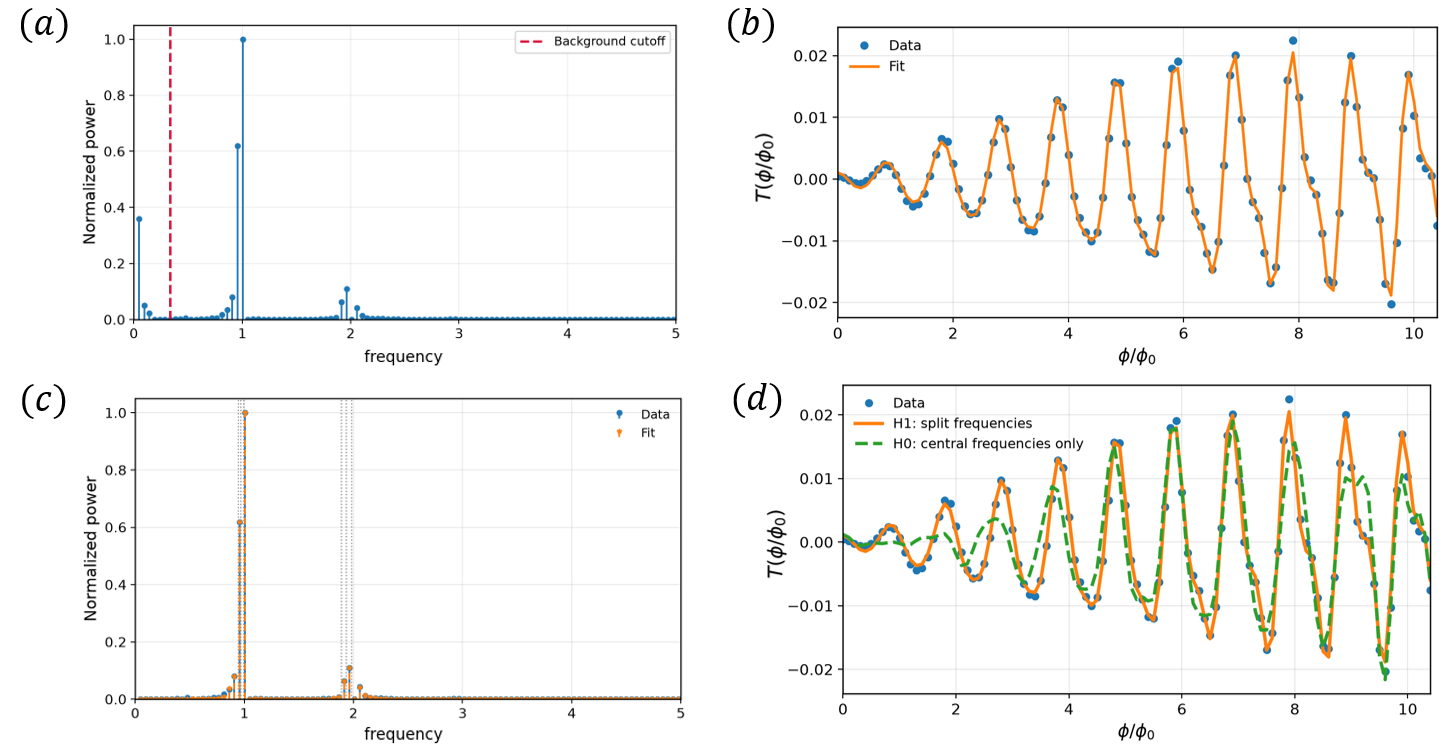}
\caption{Frequency-resolved analysis of the AB oscillations. (a) Normalized DFT power of the transmission, with the red dashed line marking the cutoff used to separate the slowly varying background (components below the cutoff) from the oscillatory signal (components above the cutoff). (b) Background-subtracted transmission and the H1 fit containing the frequency triplet and its second harmonics. (c) Comparison of the normalized DFT power of the numerical and fitted signals; vertical dotted lines mark the fitted fundamental frequencies and their second harmonics. (d) Background-subtracted transmission compared with the split-frequency model H1 and with the optimized single-frequency model H0, which contains only $f_c$ and $2f_c$ with high-order even amplitude envelopes.}
\label{fig:ab_fourier}
\end{figure*}

The relative phase can also be tuned electrostatically by applying an onsite potential $V_g$ to the green region on the upper arm. Figure~\ref{fig:ab}(c) shows the transmission as a joint function of $\Phi/\Phi_0$ and $V_g$. The alternating fringes demonstrate that either control parameter shifts the interference condition. Taking the derivative along the flux direction suppresses the smooth background and makes the fringe slopes more visible [Fig.~\ref{fig:ab}(d)]. The normalized flux is evaluated using an enclosed area $S\approx85\,\mathrm{nm}^2$, estimated as the area between the two arms.

To determine how the junction opening affects this response, the transmission is calculated as a function of magnetic flux for $W_{\mathrm{slit}}=5$, 10, 15, and $20\,\text{\AA}$. Figure~\ref{fig:ab}(e) shows that widening the slit increases the overall transmission. For every width, the equality of the positive- and negative-flux traces is consistent with the Onsager symmetry of a two-terminal device; an explicit verification of the underlying Onsager--Casimir reciprocity relation over the full flux interval is provided in the Supplemental Material [Fig.~S7(b)].

The evolution of the oscillations can be rationalized qualitatively in terms of the cyclotron radius $r_c$, which decreases with increasing magnetic field and, for a fixed area, with increasing flux. In this picture, a cyclotron orbit much larger than the slit width would hinder the formation of well-defined interferometer arms, resulting in weak or absent oscillations. Extrapolating toward higher fields, $r_c$ would eventually become comparable to or smaller than the slit width, favoring the formation of the two arms and the emergence of oscillations. Figure~\ref{fig:ab}(f) shows the increasing oscillatory modulation as $r_c/W_{\mathrm{slit}}$ decreases within the calculated range. The proposed crossover at $r_c/W_{\mathrm{slit}}\sim1$, however, lies beyond that range and represents a qualitative extrapolation rather than a
regime directly demonstrated by the present calculations.

As the first approach to the most general quantification of the oscillation in Fig.~\ref{fig:ab}(b), the following phenomenological form was adopted,
\begin{align}
T\!\left(\frac{\Phi}{\Phi_0}\right)
={}& b_0 + b_1\left|\frac{\Phi}{\Phi_0}\right|
\nonumber\\
&+ \left[ A_0 + A_1\left|\frac{\Phi}{\Phi_0}\right| \right]
\cos\!\left[ 2\pi f\left|\frac{\Phi}{\Phi_0}\right| + \varphi \right],
\label{eq:transmissionfit}
\end{align}
with fitting parameters $\lbrace b_0, b_1, A_0, A_1, f, \varphi \rbrace$. The right hand side of Eq.~(\ref{eq:transmissionfit}) consists of a first-order polynomial background and a single-frequency cosine oscillation of linearly varying envelope and it is the simplest mathematical formula to capture the AB oscillation with varying amplitude superimposed on a slowly varying non-AB background. The AB phase acquired by an electron encircling the enclosed flux is
\begin{equation}
\Delta\varphi = \frac{e}{\hbar}\oint \mathbf A\cdot d\mathbf r = 2\pi\frac{\Phi}{\Phi_0},
\label{eq:abperiod}
\end{equation}
where $\Phi=BS$ is the magnetic flux enclosed by the interferometer. The expected Aharonov--Bohm oscillation period is obtained from the condition $\Delta\varphi=2\pi$, which yields
\begin{equation}
\Delta\!\left(\frac{\Phi}{\Phi_0}\right)=1.
\label{eq:period1}
\end{equation}
Comparing this with Eq.~(\ref{eq:transmissionfit}) leads to a normalized frequency $f=1$, which is the value one should expect if the AB interference around the assumed surface indeed happens as predicted.

The dominant oscillation frequency is extracted at several energies and the fit quality is evaluated using the coefficient of determination
\begin{equation}
R^2 = 1 - \frac{\sum_{n}\left(T_n - T_n^{\mathrm{fit}}\right)^{2}}{\sum_{n}\left(T_n - \bar{T}\right)^{2}},
\label{eq:r2}
\end{equation}
where $T_n$ is the calculated transmission at the $n$th flux value, $T_n^{\mathrm{fit}}$ the corresponding value of the fit, and $\bar{T}$ the mean transmission over the flux interval; $R^2=1$ indicates a perfect fit. In Fig.~\ref{fig:ab_freq_energy}, the point at $E=2.7\,\mathrm{eV}$, pointed by arrows, lies close to $f=1$ (shown as the dashed red circle) and retains a high $R^2$ for all four slit widths, while the values of the fitted frequency lie away from the $f=1$ line for almost all other energies. This behavior is consistent with an $h/e$ AB period, as observed in monolayer and bilayer graphene-ring experiments \cite{ref27,ref28,refIwakiriBLGRing}, and with the strong directional propagation expected at the energy of maximum trigonal warping. Consistently, the oscillations are strongly suppressed already at $E\simeq2.6\,\mathrm{eV}$, where the anisotropy is weaker [SM, Fig.~S7(a)]. Thus the direct link between the AB oscillation at specific energy, the surface and geometry of the device, and the trigonal warping as the underlying mechanism, is confirmed.

The fit of Eq.~(\ref{eq:transmissionfit}) treats the flux dependence as a continuous function, whereas the numerical transmission is sampled at discrete, uniformly spaced values of $\Phi/\Phi_0$. The fit is therefore complemented with a discrete Fourier transform (DFT). The spectrum in Fig.~\ref{fig:ab_fourier}(a) contains a low-frequency contribution, a pronounced group of components around $f=1$, and a weaker group around $f=2$. The low-frequency weight arises mainly from the slowly varying transmission background, while the groups near $f=1$ and $f=2$ are consistent with the fundamental AB oscillation and its second harmonic, respectively. The observation of $h/e$ and integer higher harmonics in a gate-defined BLG ring supports this assignment \cite{refIwakiriBLGRing}; it does not, however, account for the three closely spaced components around the fundamental in the present device. This Fourier analysis allows a simple frequency-filtering separation procedure to be defined of the non-oscillatory background and the oscillatory signal, which is described in the SM (Fig.~S8).

At this point, the details of the signal variation should be investigated. Looking at Fig.~\ref{fig:ab}(b), one can observe that the oscillations behave very much cosine-like for smaller $\left|\Phi/\Phi_0\right|$, but as the flux increases, the additional dynamic becomes more pronounced, which seems to be separate from the change in background and in the oscillation amplitude. Motivated by the three closely spaced components around $f=1$, a fitting model is proposed for the background-subtracted oscillation, which is based on three oscillation frequencies: a central one $f_c$ and two symmetric side frequencies,
\begin{equation}
f_m=f_c-\Delta f,
\qquad
f_p=f_c+\Delta f.
\label{eq:frequency_triplet}
\end{equation}
The corresponding second harmonics, $2f_m$, $2f_c$, and $2f_p$, are included to account for the spectral weight around $f=2$. Because the transmission is symmetric under reversal of the magnetic flux, only cosine terms with even amplitude envelopes are retained. The fitting model, denoted H1, is
\begin{equation}
T_{\mathrm{H1}}\!\left(\frac{\Phi}{\Phi_0}\right) =
\sum_{f_i} \left[ c_{i,0} + c_{i,2} \left(\frac{\Phi}{\Phi_0}\right)^2 \right]
\cos\!\left[2\pi f_i \left(\frac{\Phi}{\Phi_0}\right) \right],
\label{eq:frequency_triplet_fit}
\end{equation}
where $f_i\in\{f_m,f_c,f_p,2f_m,2f_c,2f_p\}$. The resulting fit is compared with the numerical data in Fig.~\ref{fig:ab_fourier}(b); the determination of the optimal $f_c$ and $\Delta f$ is described in the SM [Fig.~S8(c)]. The DFT of the fitted signal closely reproduces the numerical spectrum near both the fundamental frequency $f\simeq1$ and its second harmonic $f\simeq2$ [Fig.~\ref{fig:ab_fourier}(c)]. In particular, the model captures the relative weights of the neighboring Fourier components.

As a final test, it is verified whether the oscillations can be reproduced without any frequency splitting. To this end, a single-frequency model is considered, denoted H0, which retains only the central frequency $f_c$ and its second harmonic $2f_c$,
\begin{align}
T_{\mathrm{H0}}\!\left(\frac{\Phi}{\Phi_0}\right)
={}& A_1\!\left(\frac{\Phi}{\Phi_0}\right)
\cos\!\left[ 2\pi f_c\left(\frac{\Phi}{\Phi_0}\right) \right]
\nonumber\\
&+ A_2\!\left(\frac{\Phi}{\Phi_0}\right)
\cos\!\left[ 4\pi f_c\left(\frac{\Phi}{\Phi_0}\right) \right],
\label{eq:collapsed_model}
\end{align}
where the amplitude envelopes $A_1$ and $A_2$ are even polynomials in $\Phi/\Phi_0$ of order up to twelve. For H0 the central frequency is re-optimized independently of H1, by scanning $f_c$ and determining the linear envelope coefficients by least-squares minimization at each point. The H0 model therefore contains 14 linear amplitude coefficients and one nonlinear frequency parameter, compared with 12 linear coefficients and the two nonlinear frequency parameters $f_c$ and $\Delta f$ for H1. Consequently, the models H0 and H1 have almost equal number of degrees of freedom, notably H1 is not preferred in that regard. As shown in Fig.~\ref{fig:ab_fourier}(d), the optimized H0 model nevertheless fails to reproduce the background-subtracted numerical data, and its residual exceeds that of H1 for every slit width (SM, Figs.~S9 and S10). Agreement is recovered only when the finite splitting $\Delta f$ is retained. The modulation therefore cannot be described solely by a flux-dependent amplitude at one frequency; it results from the interference of closely spaced frequency components.

This result has a natural interpretation within the Huygens--Fresnel picture. A wave emerging from an aperture is a coherent superposition of contributions that propagate along slightly different paths and acquire different phases. Likewise, the finite width of the electronic jets, coming from the finite width of the slit and visible in Fig.~\ref{fig:ab}(a), correspond to a set of contributions encircling slightly different effective areas S and thus accumulating slightly different magnetic phases. They consequently appear as the frequencies $f_m$, $f_c$, and $f_p$, rather than as a single component. Their relative phases evolve with $\Phi/\Phi_0$ and modulate the interference pattern. Restricting the description to the single frequency $f_c$ instead forces the entire flux dependence of this beating into the amplitude envelopes, which even at high polynomial order cannot reproduce the modulation [Fig.~\ref{fig:ab_fourier}(d)]. In this sense, the finite frequency splitting is the magnetic analogue of the phase dispersion between the wavelets of conventional diffraction.

\subsection{Spin polarization}

\begin{figure*}[!tp]
\centering
\includegraphics[width=0.9\textwidth]{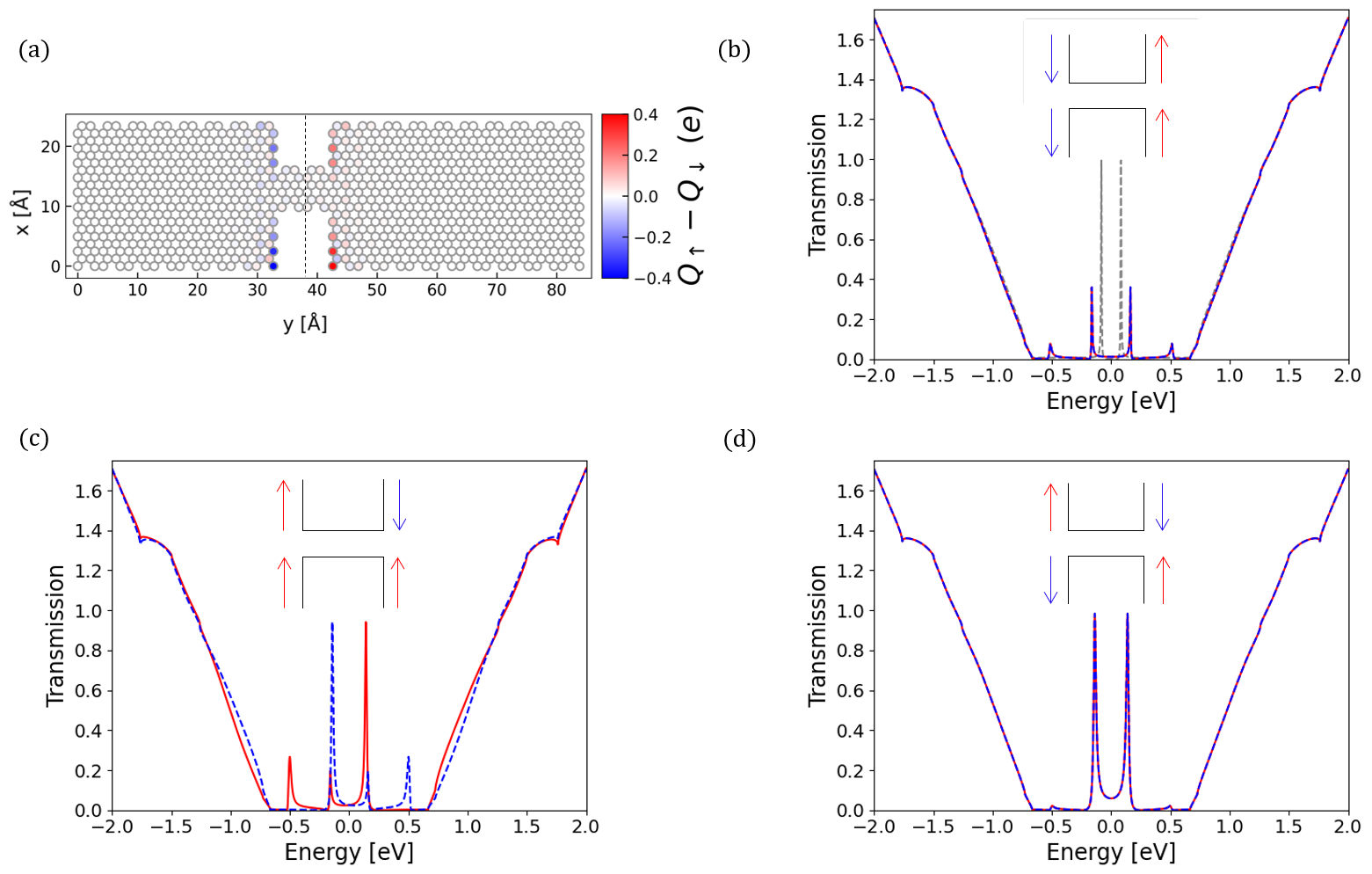}
\caption{Spin configurations and the symmetry constraint at the zigzag slit edges within the mean-field Hubbard model. In (b)--(d), solid red and dashed blue curves denote the spin-up and spin-down transmissions, respectively, and the insets show the corresponding edge-spin configurations. (a) Ground-state spin-density distribution, with its total energy taken as the reference; the black dashed line marks the left--right mirror plane. (b) Spin-resolved transmission of the antiferromagnetic ground state; the two spin channels overlap, and the grey curve shows the transmission obtained without the mean-field Hubbard term ($U=0$). (c) Spin-dependent transmission for a symmetry-breaking excited-state configuration, $15\,\mathrm{meV}$ above the ground state. (d) Unpolarized transmission for an excited-state configuration, $30\,\mathrm{meV}$ above the ground state, that preserves the combined left--right-mirror and spin-flip symmetry. Panels (b), (c), and (d) correspond to panels (a), (b), and (d) of Fig.~S11 in the SM, where the complete set of eight configurations is reported.}
\label{fig:spin_configs_main}
\end{figure*}

\begin{figure*}[!tp]
\centering
\includegraphics[width=.9\textwidth]{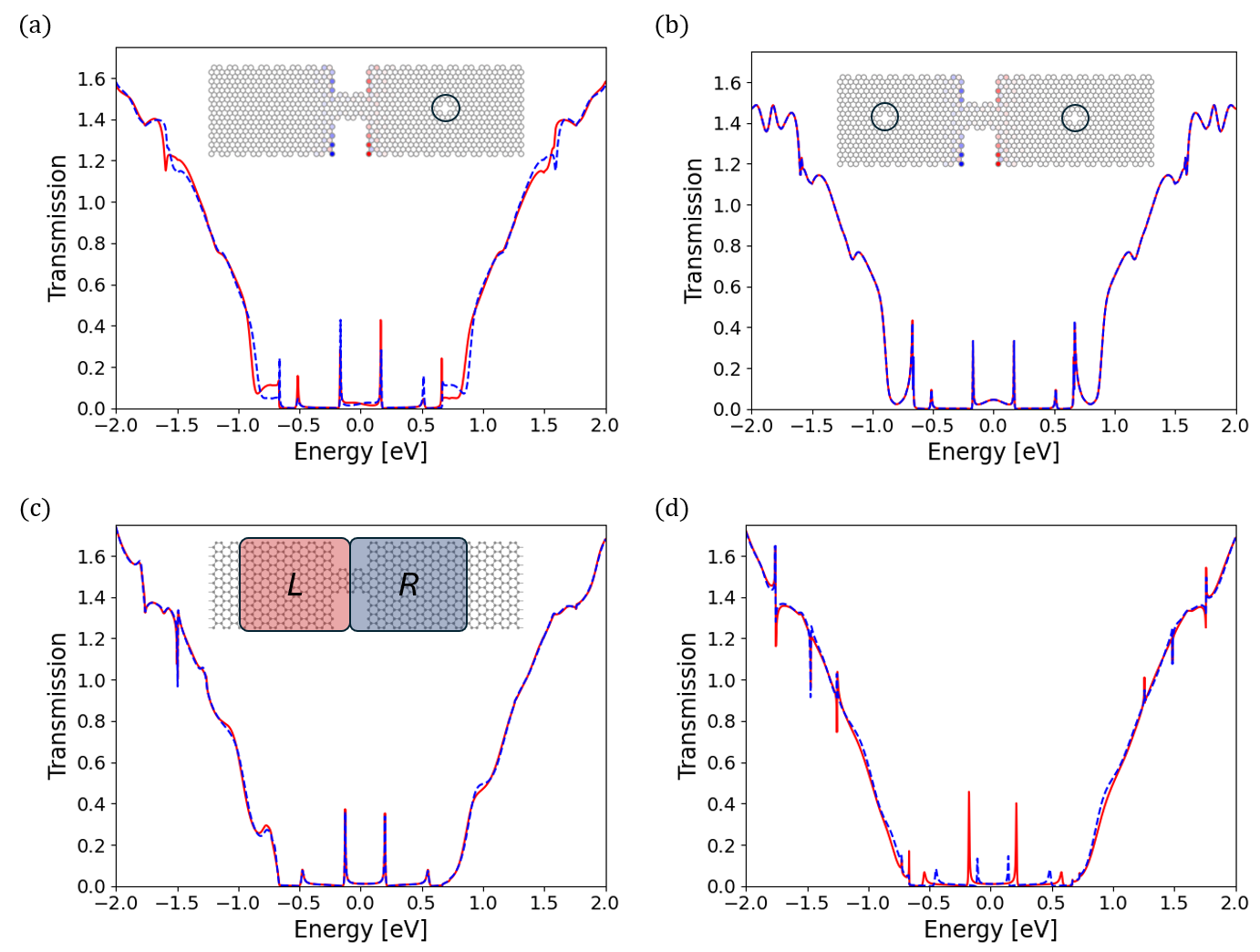}
\caption{Spin filtering induced by structural or electrostatic symmetry breaking. The solid red and dashed blue curves denote the spin-up and spin-down transmissions, respectively, and the device schematics are shown in the insets. (a) A single defect on one side of the device breaks the left--right symmetry and separates the spin channels. (b) A symmetric pair of defects restores spin-degenerate transmission. (c) A symmetric electrostatic profile, $V_L=V_R=0.1\,\mathrm{eV}$, applied to the red and blue regions of the inset, also preserves spin degeneracy. (d) An asymmetric profile, $V_L=0.1\,\mathrm{eV}$ and $V_R=0$, breaks the mirror symmetry and produces spin-dependent transmission.}
\label{fig:spin_filtering}
\end{figure*}

\begin{figure*}[!tp]
\centering
\includegraphics[width=.9\textwidth]{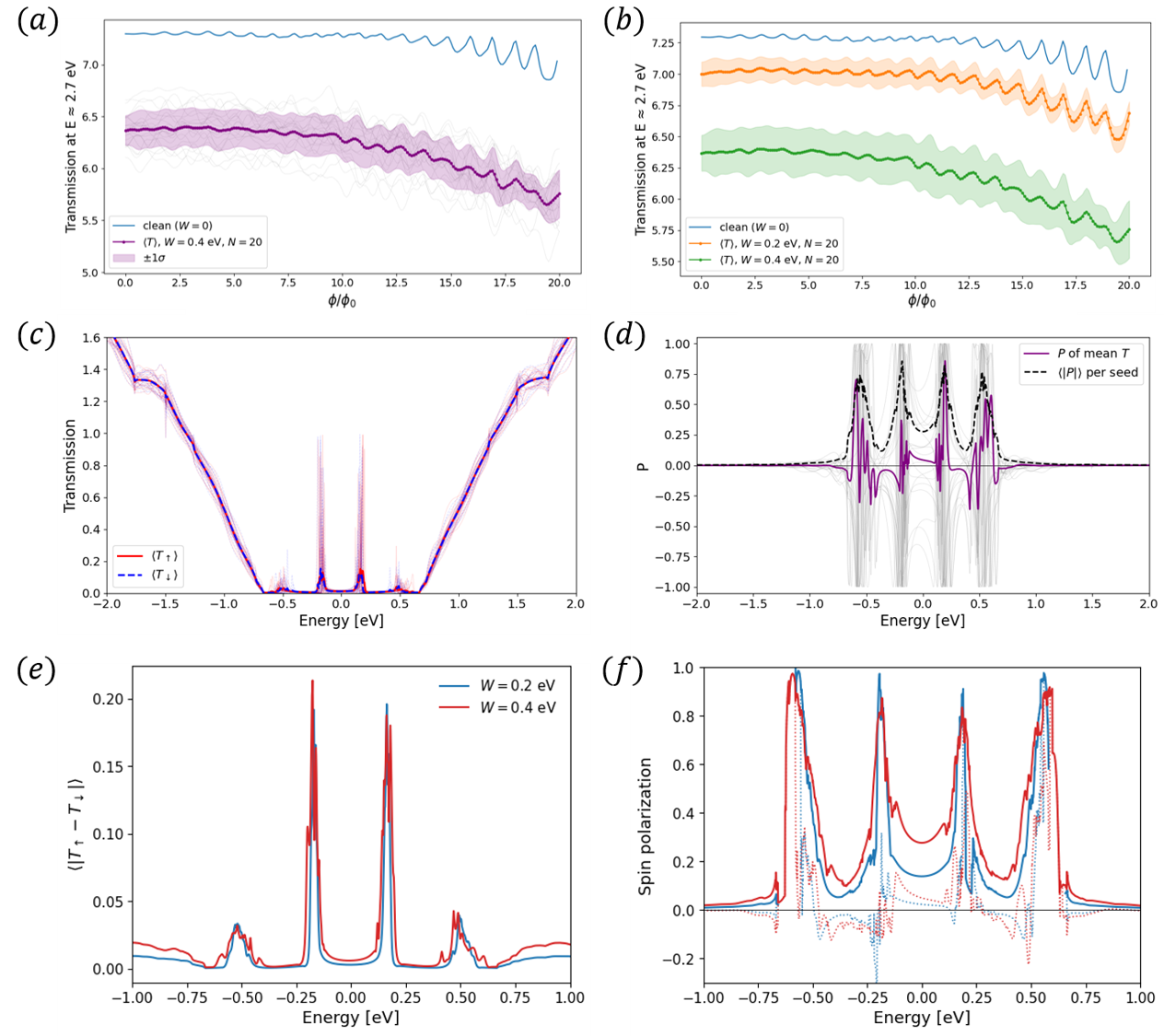}
\caption{Disorder dependence of AB interference and spin filtering for an ensemble of $N=20$ independent Anderson-disorder realizations. Panels (a) and (b) refer to the interferometer geometry of Fig.~\ref{fig:ab} with $W_{\mathrm{slit}}=20\,\text{\AA}$ at $E=2.7\,\mathrm{eV}$; panels (c)--(f) refer to the mean-field Hubbard device of Fig.~\ref{fig:spin_configs_main}, for which the self-consistent solution is converged independently for every realization. (a)~Transmission versus normalized flux for the clean device and for $W_{\mathrm{}}=0.4\,\mathrm{eV}$; faint curves are individual realizations, the solid curve is their mean, and the shaded band denotes one standard deviation. (b)~Ensemble-averaged transmission for $W_{\mathrm{}}=0.2$ and $0.4\,\mathrm{eV}$, compared with the clean result; shaded bands denote one standard deviation. (c)~Spin-resolved transmission spectra for the individual realizations and their ensemble means. (d)~Polarization formed from the ensemble-averaged transmissions, $P_{\langle T\rangle}$, compared with the ensemble mean of the realization-resolved magnitude, $\langle|P_s|\rangle$; faint curves show individual $P_s$. (e)~Ensemble-averaged spin-channel difference $\langle|T^{\uparrow}-T^{\downarrow}|\rangle$ for $W_{\mathrm{}}=0.2$ and $0.4\,\mathrm{eV}$. (f)~Corresponding energy-resolved polarization measures for the two disorder strengths; solid curves show the typical polarization magnitude and dotted curves retain its signed, sample-dependent character.}
\label{fig:disorder}
\end{figure*}

The slit necessarily introduces zigzag edge segments. Such terminations support localized edge states that develop spin-polarized configurations in interaction-based descriptions \cite{ref33}; the interplay between junction geometry and localized spin states in armchair-nanoribbon junctions has recently been analyzed systematically within the same mean-field Hubbard framework \cite{ref38}; experimental evidence for zigzag-edge magnetism has also been reported \cite{ref32}, and recent progress in atomically precise graphene nanoribbons is reviewed in Ref.~\cite{ref31}.

The edge polarization is described using the mean-field Hubbard model in Eq.~(\ref{eq:hubbard}), following its application to crossed graphene nanoribbons \cite{ref14}. The slit creates four zigzag edges, each of which can carry either spin-up or spin-down polarization. There are therefore $2^4$ nominal configurations, reduced to eight distinct configurations by global spin inversion. All eight are found to converge self-consistently; they are reported, together with their total-energy differences, in the SM, Fig.~S11. Figure~\ref{fig:spin_configs_main}(a) shows the ground-state spin-density distribution. Opposite zigzag edges align antiferromagnetically across the slit, consistent with the established behavior of zigzag graphene nanoribbons \cite{ref33}, while the two edges on the same side of the slit align ferromagnetically. The corresponding spin-resolved transmission curves overlap [Fig.~\ref{fig:spin_configs_main}(b)], so the ground state is not a spin filter.

The device has a single left--right mirror plane, shown by the black dashed line in Fig.~\ref{fig:spin_configs_main}(a). As in crossed graphene nanoribbons \cite{ref14}, this geometry imposes a useful constraint: when a left--right mirror operation followed by a spin flip leaves the Hamiltonian invariant, the two spin transmissions must be equal. This condition can be written as
\begin{equation}
\mathcal{O} H \mathcal{O}^{-1} = H \quad \Rightarrow \quad T^{\uparrow}(E) = T^{\downarrow}(E),
\label{eq:symmetry_constraint}
\end{equation}
with
\begin{equation}
\mathcal{O} = \mathcal{M}_{LR} \otimes \mathcal{F},
\label{eq:uop}
\end{equation}
where $\mathcal{M}_{LR}$ is the left--right mirror (swap) operator and $\mathcal{F}$ is the spin-flip operator, $\uparrow\rightleftharpoons\downarrow$. Exactly two of the eight configurations are invariant under $\mathcal{O}$, and these are precisely the two whose spin channels are degenerate; the remaining six all display spin-dependent transmission [SM, Fig.~S11]. Figure~\ref{fig:spin_configs_main}(d) shows one such $\mathcal{O}$-invariant excited state, which remains unpolarized. In contrast, the configuration in Fig.~\ref{fig:spin_configs_main}(c) breaks the combined symmetry and produces spin-dependent transmission. In the pristine device, this spin-filtering solution is the first excited self-consistent state, approximately $15\,\mathrm{meV}$ above the ground state. A practical spin filter should therefore break the left--right symmetry already in its ground-state either in geometry or using electrostatic profile.

This symmetry breaking is implemented in two ways. First, a defect is introduced on only one side of the device. The resulting spin-resolved transmissions are unequal [Fig.~\ref{fig:spin_filtering}(a)]. Adding a matching defect on the opposite side restores the left--right symmetry and makes the transmission spin degenerate again [Fig.~\ref{fig:spin_filtering}(b)].

Second, local electrostatic potentials $V_L$ and $V_R$ are applied to the $L$ region shown in red and the $R$ region shown in blue in the inset of Fig.~\ref{fig:spin_filtering}(c), following Eqs.~(\ref{eq:tbonsite}) and (\ref{eq:pn}). A symmetric profile, $V_L=V_R$, leaves the left--right mirror plane intact and preserves spin-degenerate transmission [Fig.~\ref{fig:spin_filtering}(c)], whereas an asymmetric profile, $V_L\neq V_R$, breaks the mirror constraint and produces spin filtering [Fig.~\ref{fig:spin_filtering}(d)]. This route provides a controllable alternative to structural defects.

\subsection{Effect of disorder}

Finally, the robustness of the interference and spin-filtering responses is tested against Anderson disorder. An ensemble of $N=20$ independent disorder realizations is considered. The interference data of Figs.~\ref{fig:disorder}(a) and \ref{fig:disorder}(b) were obtained for the interferometer geometry of Fig.~\ref{fig:ab} with $W_{\mathrm{slit}}=20\,\text{\AA}$ at $E=2.7\,\mathrm{eV}$. For the spin-resolved results, the mean-field Hubbard equations were solved self-consistently for each disorder realization separately.

Figure~\ref{fig:disorder}(a) compares the individual transmission traces at $W=0.4\,\mathrm{eV}$ with their ensemble mean and with the clean result. Although each realization distorts the AB oscillations differently, the mean retains a damped oscillatory component. The comparison of $W=0.2$ and $0.4\,\mathrm{eV}$ in Fig.~\ref{fig:disorder}(b) shows that the oscillation amplitude decreases as the disorder strength increases, indicating progressive loss of coherence between the two arms.

For each realization $s$, the spin polarization is defined as
\begin{equation}
P_s(E)=\frac{T_s^{\uparrow}(E)-T_s^{\downarrow}(E)}{T_s^{\uparrow}(E)+T_s^{\downarrow}(E)}.
\label{eq:spin_polarization}
\end{equation}
Figure~\ref{fig:disorder}(c) shows the spin-resolved transmissions for the individual realizations and their ensemble means. Because different disorder configurations favor opposite polarization signs, averaging $T^{\uparrow}$ and $T^{\downarrow}$ before evaluating Eq.~(\ref{eq:spin_polarization}) produces a small net polarization. By contrast, the ensemble average of the magnitude, $\langle|P_s|\rangle$, remains large and individual realizations can approach complete polarization near the resonances [Fig.~\ref{fig:disorder}(d)].

Figures~\ref{fig:disorder}(e) and \ref{fig:disorder}(f) show that, within the disorder range considered here, increasing $W$ enhances both the average spin-channel splitting and the typical polarization magnitude. Anderson disorder therefore strengthens spin filtering in a given realization by breaking the device symmetry more strongly, even though the signed polarization cancels when realizations with opposite asymmetries are averaged together.

\section{Discussion}
Several considerations determine how the mechanisms studied here translate to experiment. In monolayer graphene, maximal trigonal warping occurs at $E\simeq|t|$, far beyond electrostatically accessible doping; this energy was chosen to isolate the kinematics of directional jetting, not as a proposal for monolayer operation. The natural experimental
implementation of the charge- and valley-interferometric part of the geometry is instead a gate-defined or etched slit in BLG, where the same threefold anisotropy appears at meV-to-tens-of-meV energies, and the
four-band calculation of Appendix~\ref{app:bilayer} confirms beam formation in such a device at $E=50\,\mathrm{meV}$. Because the interference relies only on two ballistic paths enclosing a flux, and AB oscillations in gate-defined BLG rings are established both experimentally and in four-band calculations \cite{refIwakiriBLGRing,refChen}, the AB response is expected to carry over qualitatively. The BLG interference and spin-filtering responses nevertheless remain to be investigated explicitly.

Two limitations of the model deserve mention. First, the nearest-neighbor $\pi$-band Hamiltonian neglects longer-range hoppings, which introduce electron--hole asymmetry and quantitatively modify the isoenergy contours near $E=|t|$. These corrections, however, preserve the threefold symmetry of the contours around each valley, which is the only ingredient required for directional, valley-selective jetting, so the mechanism itself is robust. Second, one flux quantum through the simulated enclosed area $S\approx85\,\mathrm{nm}^2$ corresponds to a magnetic field of order $\Phi_0/S\approx50\,\mathrm{T}$; such fields serve here as a numerically convenient means of scanning the AB phase in a device of tractable size. Because the interference is controlled by the dimensionless flux $\Phi/\Phi_0$, a lithographically realistic device with an enclosed area of order $10^4\,\mathrm{nm}^2$ would exhibit the same oscillations at fields below $1\,\mathrm{T}$, where the Zeeman coupling neglected in the calculations is also negligible.

Compared with crossed graphene nanoribbons \cite{ref13,ref14,ref15}, the slit junction operates within a single flake and dispenses with the stacking and angular alignment of two separately grown ribbons. The charge- and valley-interferometric functionalities do not require atomically perfect edges: the AB oscillations persist under moderate Anderson disorder (Sec.~III\,D). The spin-filtering functionality is more demanding, since it relies on magnetically ordered zigzag segments; experimental reports of room-temperature magnetic order on zigzag edges \cite{ref32} and progress in atomically precise nanoribbon synthesis \cite{ref31} suggest that this regime is within reach, although edge passivation and reconstruction remain practical concerns. Importantly, the electrostatic route to symmetry breaking [Fig.~\ref{fig:spin_filtering}(d)] provides a deterministic, gate-tunable control knob that does not depend on the placement of structural defects.

\section{Conclusion}

In summary, the graphene slit junction combines valley-selective beam formation, coherent path recombination, and symmetry-controlled spin filtering within a single compact geometry. Trigonal warping
produces two directional electron jets that form an AB interferometer after reflection and recombination at the drain, while the resolved frequency triplet accounts for the modulation of the oscillation amplitude. A local onsite potential shifts the interference phase
independently of the magnetic flux. In a separate MFH calculation, the pristine antiferromagnetic ground state remains spin unpolarized because of its combined mirror and spin-flip symmetry. An asymmetric electrostatic
profile breaks that symmetry and yields spin-dependent transmission without structural defects, so that electrostatic gating controls both the interference phase and the spin response. Anderson disorder damps the AB oscillations and induces a spin polarization whose sign varies from sample to sample. Although the atomistic calculations use monolayer graphene, the
shared threefold group-velocity anisotropy motivates a low-energy BLG implementation of the interferometer.

\section*{Data Availability}

The data that support the findings of this article are openly
available in Ref.~\cite{zenodo_data}.

\begin{acknowledgments}
OCP Foundation has supported this work with project grant AS70, ``Towards phosphorene-based materials and devices,'' and with the support of the Chair ``Multiphysics and HPC'' led by Mohammed VI Polytechnic University. The authors gratefully acknowledge the support and computing resources from the Toubkal supercomputer \cite{ref34} at Mohammed VI Polytechnic University (Morocco).
\end{acknowledgments}

\section*{Conflict of interest}
The authors declare no competing interests.

\appendix

\section{Bilayer graphene: models, energy scales, and finite-size limitations}
\label{app:bilayer}

\begin{figure*}[!tp]
    \centering
    \includegraphics[width=\textwidth]{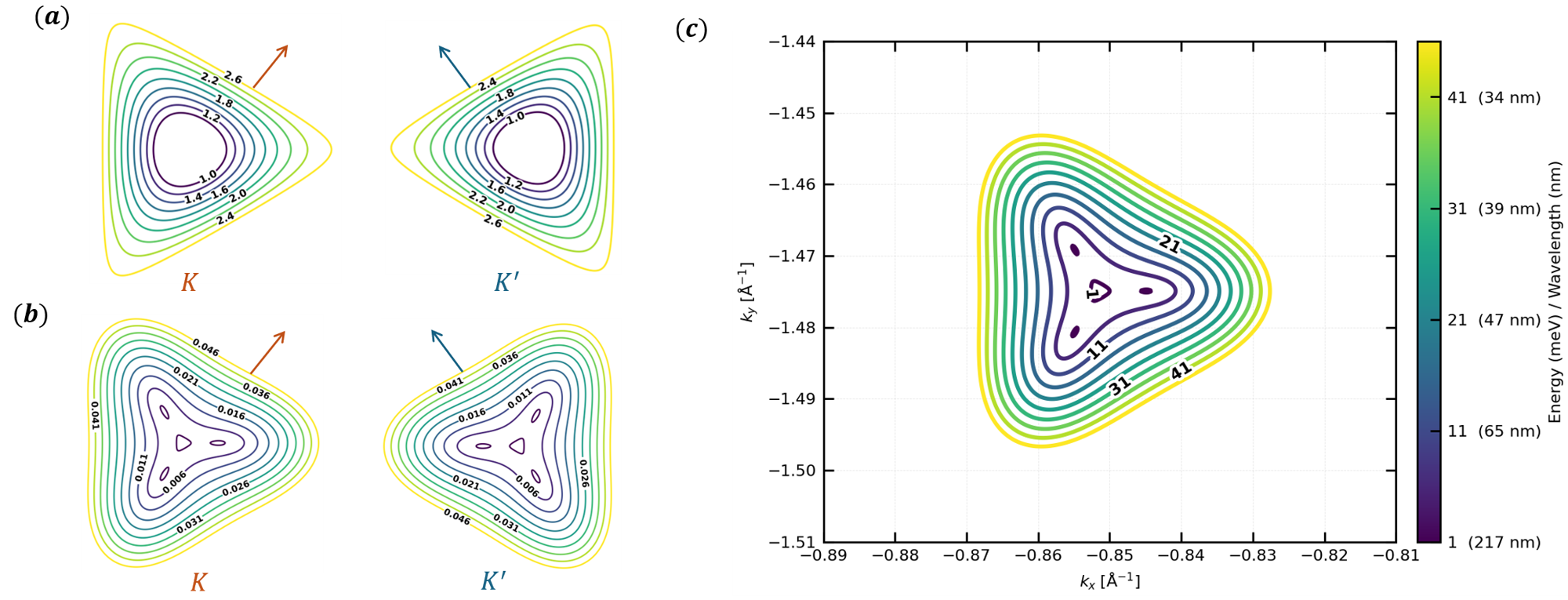}
    \caption{Trigonal warping and the associated energy and length scales in monolayer and bilayer graphene. Isoenergy contours (labels in eV) around the K and K$^\prime$ valleys of (a) monolayer graphene and (b) bilayer graphene; the arrows indicate the corresponding group-velocity directions. The two systems display qualitatively analogous threefold group-velocity anisotropy, at electron-volt energies in the monolayer and meV-to-tens-of-meV energies in the bilayer. (c) Zoom on a single bilayer valley, with isoenergy contours labeled in meV. The color bar also reports the estimated characteristic Fermi wavelengths $\lambda_F$.}
    \label{fig:bilayer_contours}
\end{figure*}

\begin{figure*}[!tp]
    \centering
    \includegraphics[width=\textwidth]{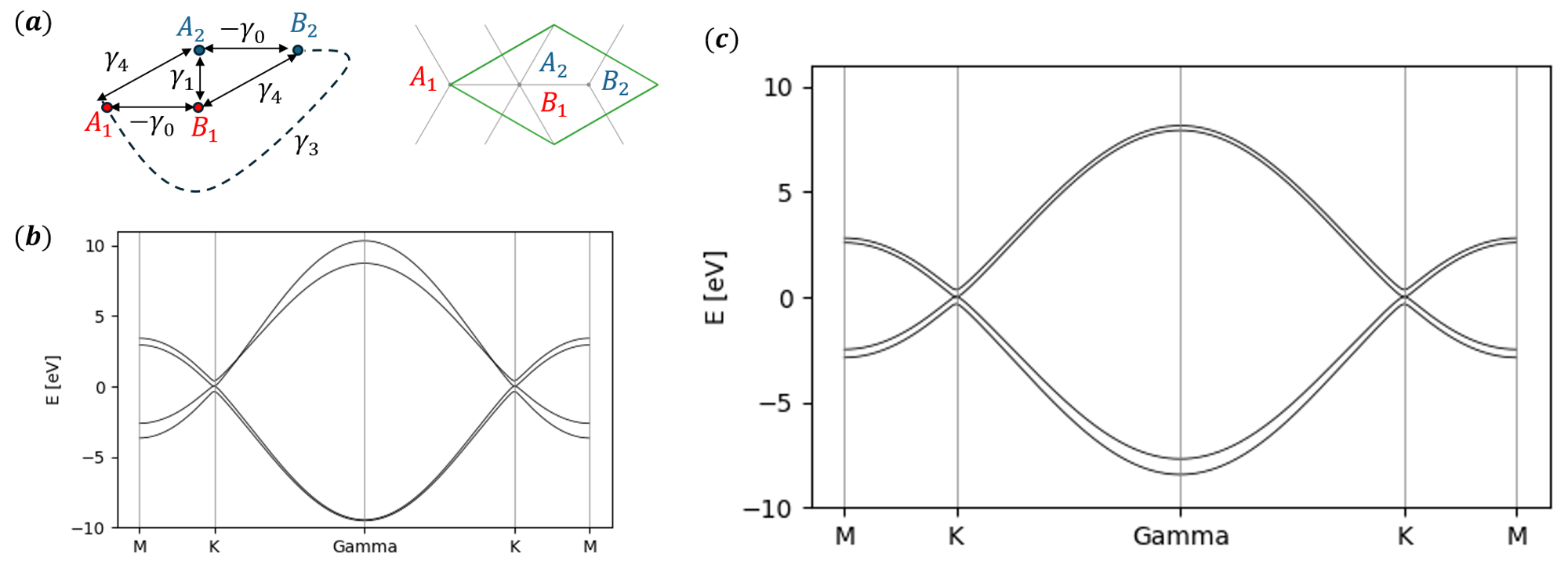}
    \caption{Atomistic models of Bernal-stacked bilayer graphene. (a) Hopping parameters of the Slonczewski--Weiss--McClure (SWM) model \cite{refSlonczewski,refMcClure,refMcCann} (left) and the corresponding real-space unit cell (right), with the $A_1,B_1$ and $A_2,B_2$ sublattices of the bottom and top layers. (b) Band structure along M--K--$\Gamma$--K--M obtained with the SWM parametrization. (c) Band structure along the same path obtained with the Slater--Koster parametrization used in the crossed-ribbon literature \cite{ref13,ref14,ref15}. The two descriptions differ quantitatively in the bandwidth at $\Gamma$ and in the interlayer splitting, but share the same low-energy structure near K and yield qualitatively similar transport.}
    \label{fig:bilayer_models}
\end{figure*}

Figure~\ref{fig:bilayer_contours} compares isoenergy contours in monolayer graphene and BLG. Both display the qualitatively analogous threefold group-velocity anisotropy responsible for valley-dependent jetting, but at electron-volt energies in the monolayer, with the strongest warping near $E=2.7\,\mathrm{eV}$, and at meV-to-tens-of-meV energies in BLG, within reach of electrostatic gating.

Figure~\ref{fig:bilayer_contours}(c) makes the length-scale consequence explicit: the representative Fermi wavelength $\lambda_F$ grows
from $\simeq34\,\mathrm{nm}$ at $41\,\mathrm{meV}$ to $\simeq217\,\mathrm{nm}$ at $1\,\mathrm{meV}$. Since detailed atomistic description is computationally constricted by system size, this comparison supports the use of monolayer graphene in the main text as a mechanism-level model, while bilayer graphene provides the experimentally realistic low-energy platform.
\begin{figure*}[!tp]
    \centering
    \includegraphics[width=\textwidth]{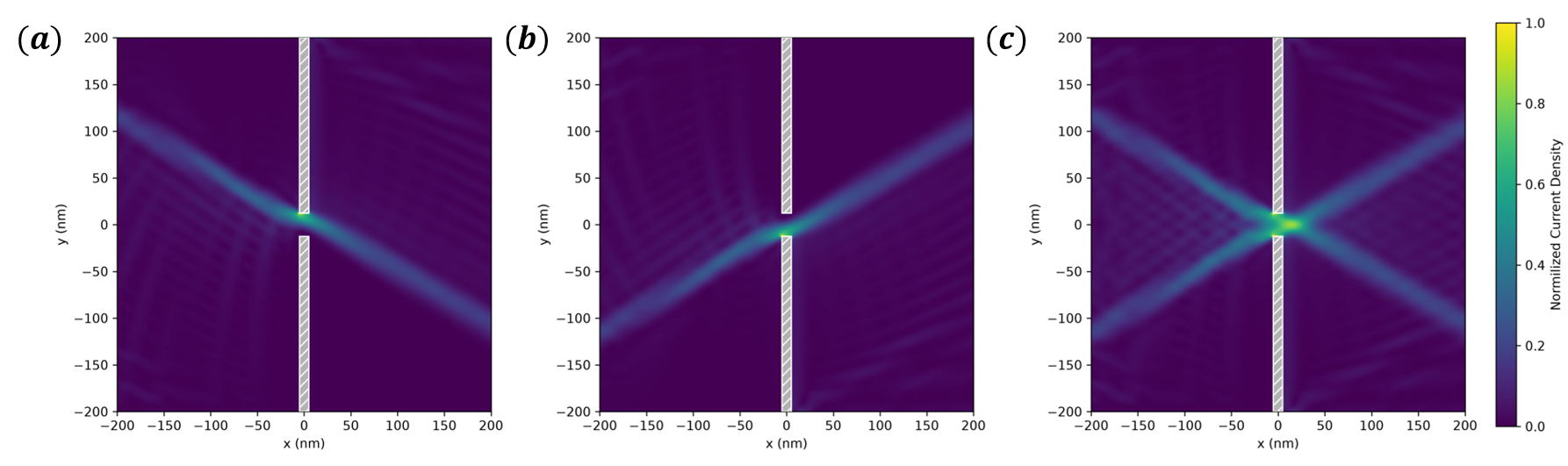}
    \caption{Low-energy beam formation in a BLG slit device described by a four-band continuum Hamiltonian discretized in Kwant. Normalized current-density magnitude for the highest-transmission eigenchannel of the (a) K and (b) K$^\prime$ valleys, and (c) their vector-current sum, at $E=50\,\mathrm{meV}$ and zero magnetic field. The nominal $400\times400\,\mathrm{nm}^2$ scattering region has a $25\,\mathrm{nm}$ opening through a $10\,\mathrm{nm}$-thick etched barrier (hatched regions) and is connected to matching semi-infinite leads on the left and right. Implementation details and control calculations are given in the Supplemental Material.}
    \label{fig:blg_slit_check}
\end{figure*}
To model BLG explicitly, the Slonczewski--Weiss--McClure (SWM) model \cite{refSlonczewski,refMcClure,refMcCann} is  implemented. The hopping magnitudes were $(\gamma_0,\gamma_1,\gamma_3,\gamma_4)=(3.16,0.381,0.38,0.14)\,\mathrm{eV}$, the central values extracted from infrared spectroscopy in Ref.~\cite{refKuzmenko}. Its hopping structure and real-space unit cell are sketched in Fig.~\ref{fig:bilayer_models}(a), and the calculated band structure in Fig.~\ref{fig:bilayer_models}(b) reproduces the expected bilayer dispersion. To test the robustness of these conclusions with respect to the parametrization, the Slater--Koster form of Ref.~\cite{ref13} is  also implemented, using its fitted values $t_{\parallel}=2.682\,\mathrm{eV}$, $t_{\perp}=0.371\,\mathrm{eV}$, the in-plane next-nearest-neighbor hopping $t^\prime=2.7\,\mathrm{meV}$, carbon--carbon distance $a=1.42\,\text{\AA}$, and interlayer distance $d=3.34\,\text{\AA}$; the same parametrization is used in Refs.~\cite{ref14,ref15}. The corresponding band structure is shown in Fig.~\ref{fig:bilayer_models}(c). Although the details of the dispersion change, the transport behavior remains qualitatively similar. In both cases, the simulated atomistic bilayer slit device does not recover the degree of valley polarization inferred for the experimental devices.

A likely reason is a mismatch of length scales, quantified in Fig.~\ref{fig:bilayer_contours}(c). Experimental gate-defined bilayer constrictions have widths of order $50\,\mathrm{nm}$ \cite{refInglaTEF}, whereas the present atomistic simulations are limited to widths of only a few nanometers. In the energy window where the bilayer contours are strongly warped, the representative Fermi wavelength is therefore comparable to, or larger than, the simulated device width, which strongly suppresses the clean jetting regime. In the geometry studied in Ref.~\cite{ref37}, the preferred propagation directions became well resolved only when the simulated ballistic region spanned several Fermi wavelengths. A quantitative slit design should therefore scale both the aperture width and the propagation region with the relevant directional wavelength, rather than comparing the slit width alone with experimental dimensions.

Published BLG transport calculations reveal a consistent tradeoff between atomic resolution and experimentally relevant device dimensions. The transverse-periodic real-space Green-function scheme developed for bulk monolayer-graphene heterojunctions \cite{refLiuBulkMLG} was applied to native-carbon Bernal-BLG junctions using simplified Hamiltonians containing $\gamma_0$, $\gamma_1$, and layer-dependent onsite potentials, but not the skew hoppings $\gamma_3$ and $\gamma_4$ \cite{refLiuBLGFP}. Larger BLG structures were treated with scaled honeycomb lattices \cite{ref41}: Du \emph{et al.} used a scaling factor $s_f=4$ for a laterally invariant junction \cite{refLiuAntiKlein}, and Schrepfer \emph{et al.} used $s_f=4$ for a two-dimensional cavity, without specifying explicit $\gamma_3$ or $\gamma_4$ terms in either calculation \cite{ref39}. Native-carbon finite-device calculations have instead focused on nanoscale or high-field geometries, including a hybrid monolayer--bilayer Hall interferometer retaining only $\gamma_0$ and $\gamma_1$ \cite{refMirzakhaniMLGBLG}.

Focusing on the atomistic BLG simulations, a four-hopping carbon-lattice model including skew interlayer coupling was used to determine the Fermi contour and injection distribution of a reduced quantum-point-contact region before semiclassical propagation through the complete focusing device \cite{refInglaTEF}. It does not, however, provide a native-carbon coherent-transport solution for the complete experimentally sized device. The gate-defined BLG ring experiment likewise employed electrostatic finite-element and semiclassical analyses rather than a microscopic electronic-transport calculation \cite{refIwakiriBLGRing}.

\begin{figure*}[!tp]
    \centering
    \includegraphics[width=\textwidth]{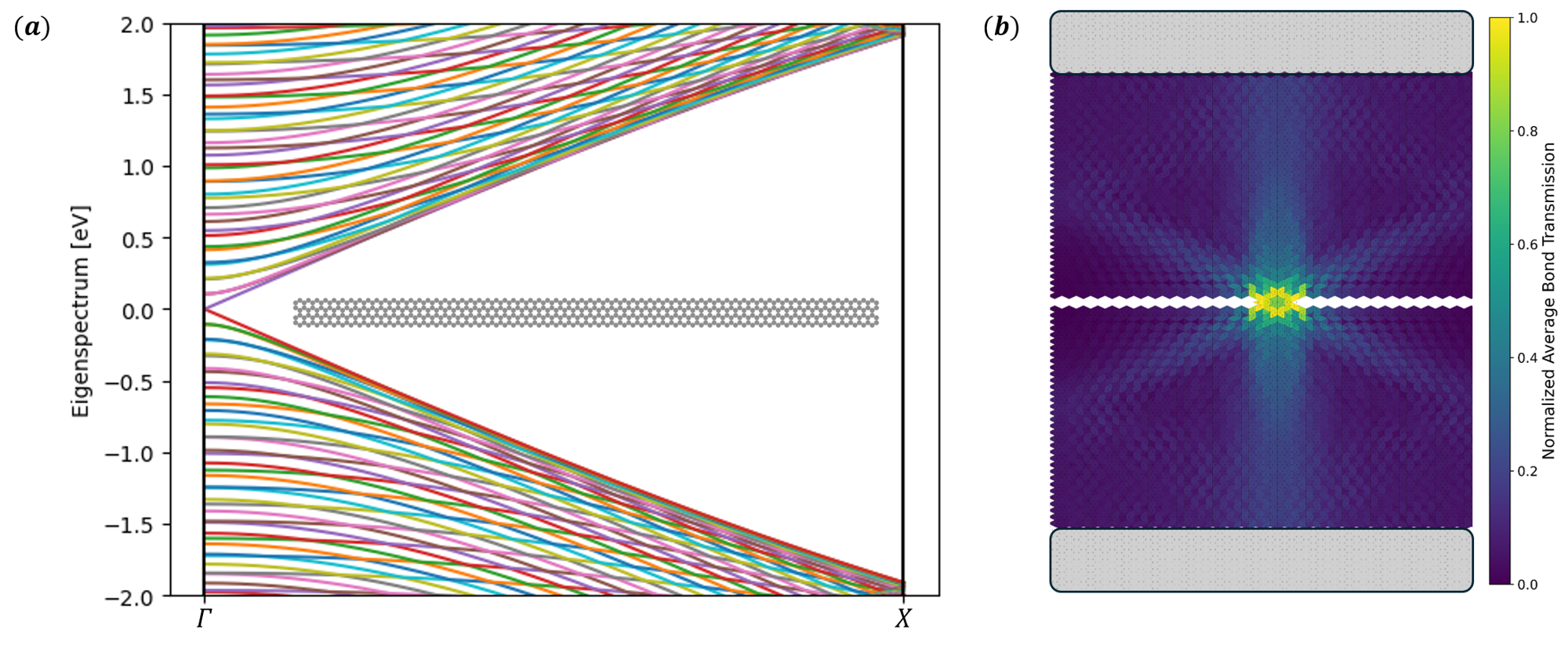}
    \caption{Electrode characterization and anisotropic propagation in the zigzag device. (a) Band structure of the armchair graphene nanoribbon electrode used for the calculations in the main text, demonstrating metallic behavior; the inset shows the corresponding electrode lattice. (b) Bond-transmission map of the zigzag graphene nanoribbon device at $E = 2\,\mathrm{eV}$, revealing three distinct anisotropic propagation directions on each side of the slit, consistent with the trigonal-warping picture discussed in the main text.}
    \label{fig:jets_device}
\end{figure*}

For realistic extended two-dimensional devices, Chen \emph{et al.} discretized a four-band continuum Hamiltonian on an auxiliary square grid \cite{refChen}. Their model retained $\gamma_1$ but omitted $\gamma_3$, and consequently did not reproduce low-energy trigonal warping. The quantum ring was implemented in Kwant~\cite{refKwant}, whereas their other transport examples used wave-function matching. Schoeppl \emph{et al.} retained the square-grid four-band strategy with a spacing of $1\,\mathrm{nm}$, restored $\gamma_3$ to describe the anisotropic Fermi contour, and used Kwant-based coherent scattering calculations together with semiclassical equations of motion \cite{ref37}. That work omitted $\gamma_4$ and reported an approximately nineteen-fold reduction of the scattering-matrix size relative to an atomistic lattice.

To test the proposed connection to BLG directly, the same four-band continuum model and square-grid discretization in Kwant~\cite{ref37} was implemented, while retaining the slit geometry rather than the electrostatically confined cavity of that work. At an injection energy of $E=50\,\mathrm{meV}$ and
crystal orientation $\alpha=0^\circ$, the current-density
maps of the highest-transmission eigenchannel in each valley show distinct beams on opposite sides of the transport axis [Fig.~\ref{fig:blg_slit_check}(a,b)] separated by $60^\circ$, similar to the case in monolayer graphene. Their valley sum recovers a two-beam pattern downstream of the slit [Fig.~\ref{fig:blg_slit_check}(c)], providing a direct, model-level check that valley-dependent beam formation also occurs in BLG at a much lower energy than in the monolayer calculations. Calculations with $\gamma_3=0$ and with the crystal rotated by $90^\circ$ relative to the slit are presented in the Supplemental Material, together with the Hamiltonian, device parameters and discretization. This comparison concerns zero-field beam formation within a valley-conserving continuum model; it does not establish robustness against atomic-edge intervalley scattering or constitute a BLG calculation of the AB or spin-filtering response.

\section{Electrode characterization and multi-electrode detection of interference}
\label{app:jets}

\begin{figure*}[!tp]
    \centering
    \includegraphics[width=.8\textwidth]{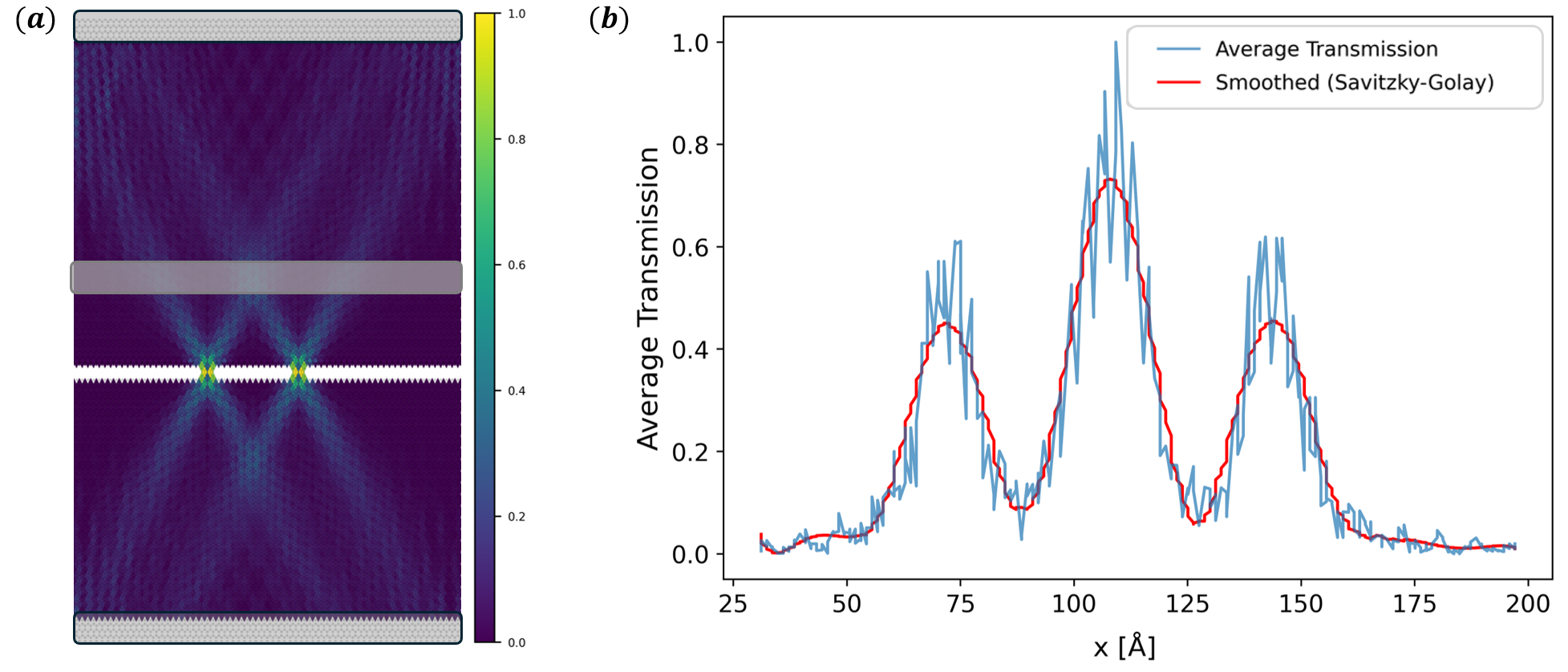}
    \caption{Interference-pattern analysis in a double-slit graphene junction. (a) Bond-transmission map at $E = 2\,\mathrm{eV}$ with the detection region highlighted in the central gray area, analogous to a detection screen in optics. (b) Spatially averaged transmission across the detection region as a function of the transverse coordinate $x$ (in \AA), shown as the raw profile together with a Savitzky--Golay smoothing. The two side peaks correspond to the independently propagating beams and the central peak to their constructive interference.}
    \label{fig:double_slit_detection}
\end{figure*}

\begin{figure*}[!tp]
    \centering
    \includegraphics[width=.8\textwidth]{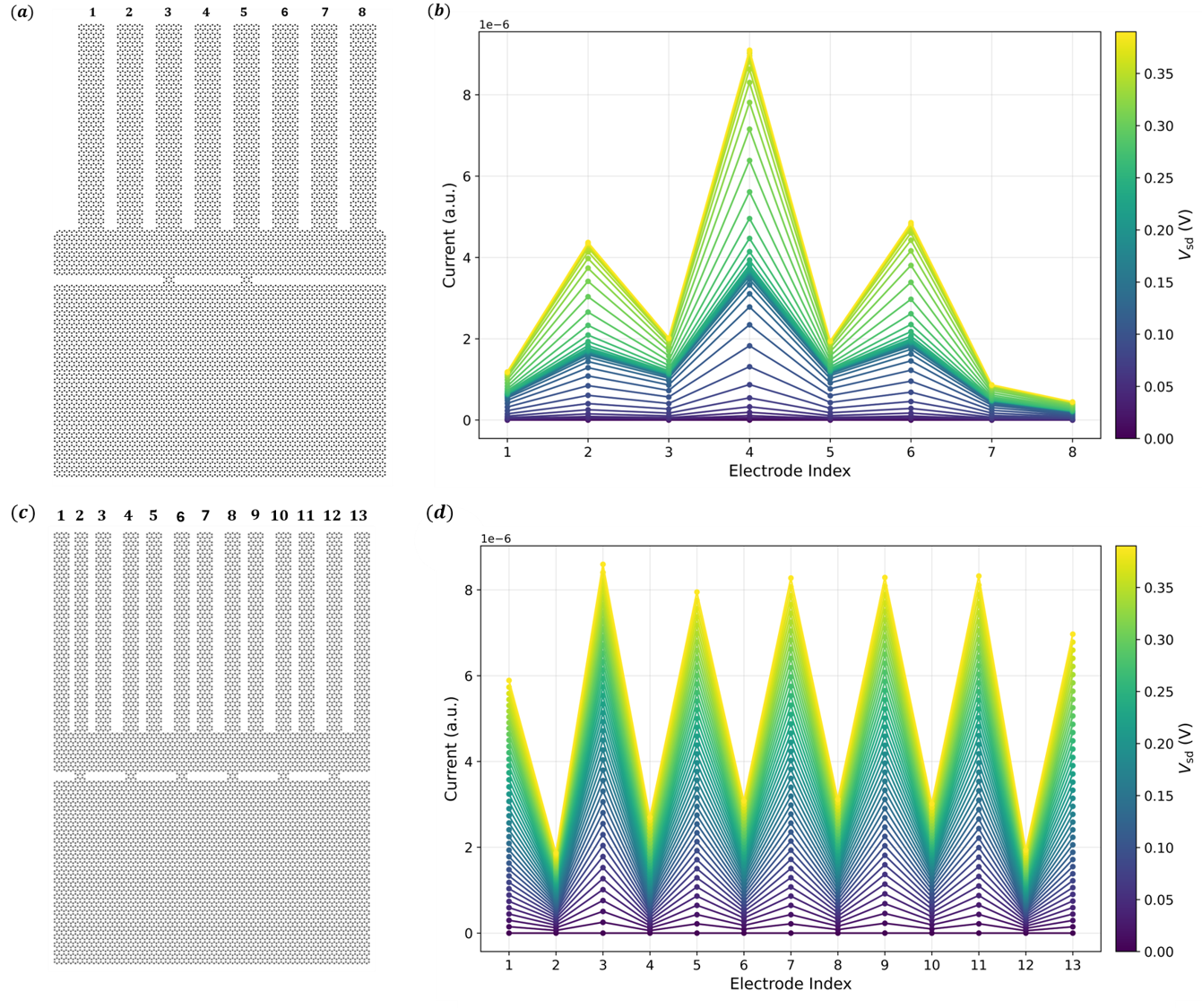}
    \caption{Multi-electrode detection of interference and diffraction. (a) Geometry of the double-slit graphene nanoribbon coupled to an array of eight detector electrodes. (b) Calculated currents at the individual electrodes for a range of source--drain voltages, reproducing the three-peak profile of Fig.~\ref{fig:double_slit_detection}(b). (c) Geometry of the multi-slit device coupled to a thirteen-electrode array. (d) Corresponding current distribution, which displays a diffraction-grating-like pattern. All currents follow from Eqs.~(\ref{eq:current}) and (\ref{eq:fermi}) under the assumption of bias-independent transmission.}
    \label{fig:multielectrode_detection}
\end{figure*}

To model the armchair device discussed in the main text, the electrode configuration shown in the inset of Fig.~\ref{fig:jets_device}(a) is employed. The corresponding band structure confirms metallic behavior, ensuring efficient carrier injection. For the zigzag geometry, the bond-transmission map in Fig.~\ref{fig:jets_device}(b) reveals three distinct anisotropic propagation directions, consistent with the trigonal-warping picture discussed in the main text.

To analyze the interference pattern generated by two slit junctions, a detection region is defined where the two beams overlap [central gray area in Fig.~\ref{fig:double_slit_detection}(a)], and the spatially averaged transmission within that region is computed. The resulting profile, plotted as a function of the transverse coordinate $x$ in Fig.~\ref{fig:double_slit_detection}(b), exhibits three peaks: two side peaks associated with the independent outgoing beams and a central peak associated with constructive interference.

This interference pattern could be probed experimentally using a multi-electrode geometry analogous to that employed in recent graphene electron-optics measurements \cite{ref35}. In the device sketched in Fig.~\ref{fig:multielectrode_detection}(a), the interference profile is reconstructed from the current distribution over eight detector electrodes. The currents shown in Fig.~\ref{fig:multielectrode_detection}(b) were calculated from the Landauer--B\"uttiker formalism of Eqs.~(\ref{eq:current}) and (\ref{eq:fermi}), under the assumption of bias-independent transmission, and reproduce the three-peak profile obtained from the bond-transmission map. The same approach extends naturally to multiple-slit devices: Figure~\ref{fig:multielectrode_detection}(c) shows a graphene nanoribbon with several slits coupled to a thirteen-electrode array, and the corresponding current distribution [Fig.~\ref{fig:multielectrode_detection}(d)] displays a diffraction-grating-like pattern.

\clearpage
\bibliography{refs}

\clearpage
\setcounter{figure}{0}
\setcounter{table}{0}
\setcounter{equation}{0}
\renewcommand{\thefigure}{S\arabic{figure}}
\renewcommand{\thetable}{S\arabic{table}}
\renewcommand{\theequation}{S\arabic{equation}}

\section*{Supplemental Material}
\section*{Summary of Content}

This Supplemental Material contains additional numerical details,
validation tests, and extended figures supporting the main manuscript.
Section~S1 details the magnetic-field implementation and
its validation through band-structure, Hofstadter, and Hall-bar benchmarks,
followed by magnetotransport in the slit geometry.
Section~S2 provides additional checks of the Aharonov--Bohm response, including Onsager--Casimir reciprocity,
and describes the Fourier filtering and frequency-analysis procedures used to characterize the oscillations.
Section~S3 summarizes the self-consistent spin configurations, their symmetry relations, and electrostatic control of beam propagation.
Section~S4 describes the four-band continuum Hamiltonian, model parameters, and numerical regularization used for the low-energy bilayer-graphene calculations. Valley-resolved current maps illustrate the effects of suppressing trigonal warping and rotating the crystal relative to the slit.
The broader discussion of bilayer-graphene models and the multi-electrode detection scheme is provided in Appendixes~A and B
of the main text, respectively.


\section*{S1. Magnetic-Field Implementation and Validation}

\begin{figure*}[!tp]
    \centering
    \includegraphics[width=.9\textwidth]{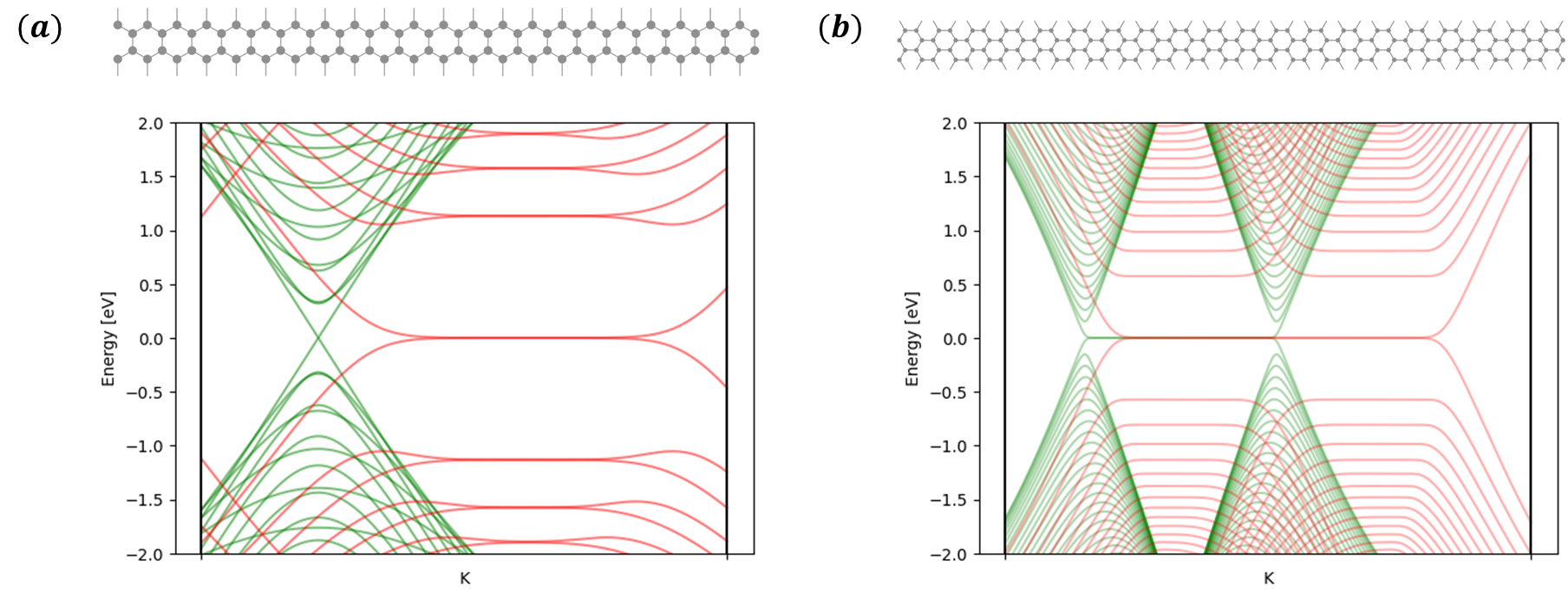}
    \caption{
    Validation of the magnetic-field implementation.
    Calculated band structures for
    (a) armchair and
    (b) zigzag graphene nanoribbons
    in the absence of magnetic field (green) and in the presence
    of a perpendicular magnetic field (red), showing the expected
    Landau-level formation.
    }
    \label{fig:magnetic_bands}
\end{figure*}

\begin{figure*}[!tp]
    \centering
    \includegraphics[width=.9\textwidth]{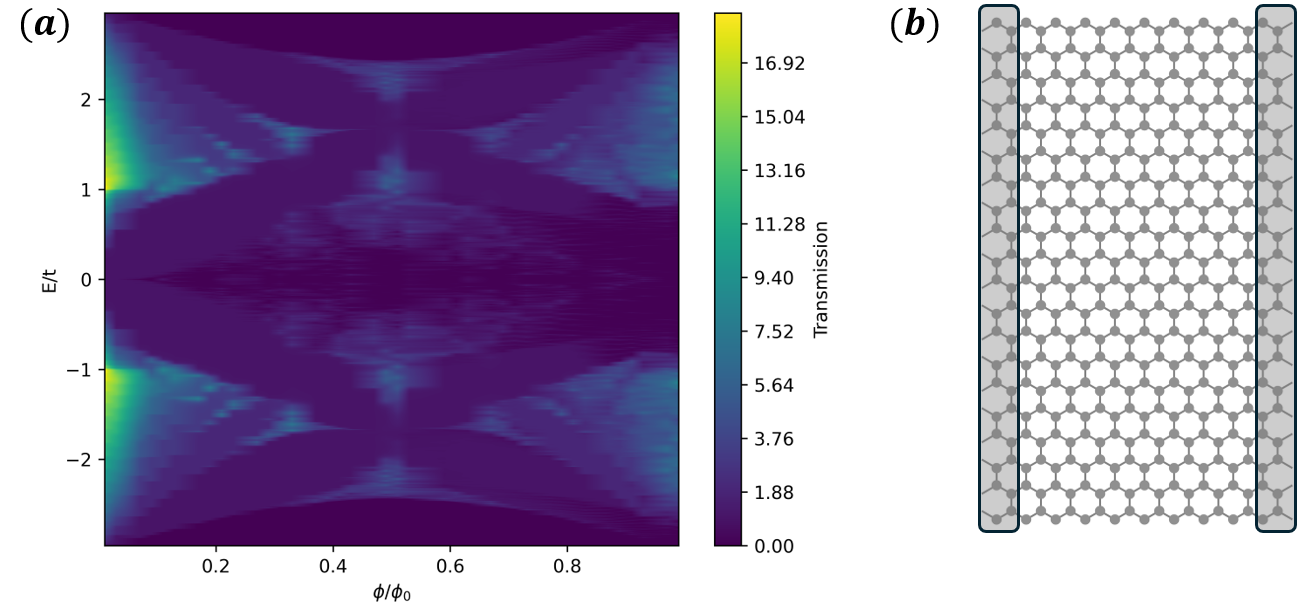}
    \caption{
    Verification of the Hofstadter butterfly spectrum.
    (a) Magnetotransport spectrum obtained from transmission
    calculations as a function of magnetic flux and energy.
    (b) Geometry of the zigzag graphene nanoribbon used in the
    simulation, with leads indicated in gray.
    }
    \label{fig:hofstadter}
\end{figure*}

\begin{figure*}[!tp]
    \centering
    \includegraphics[width=.9\textwidth]{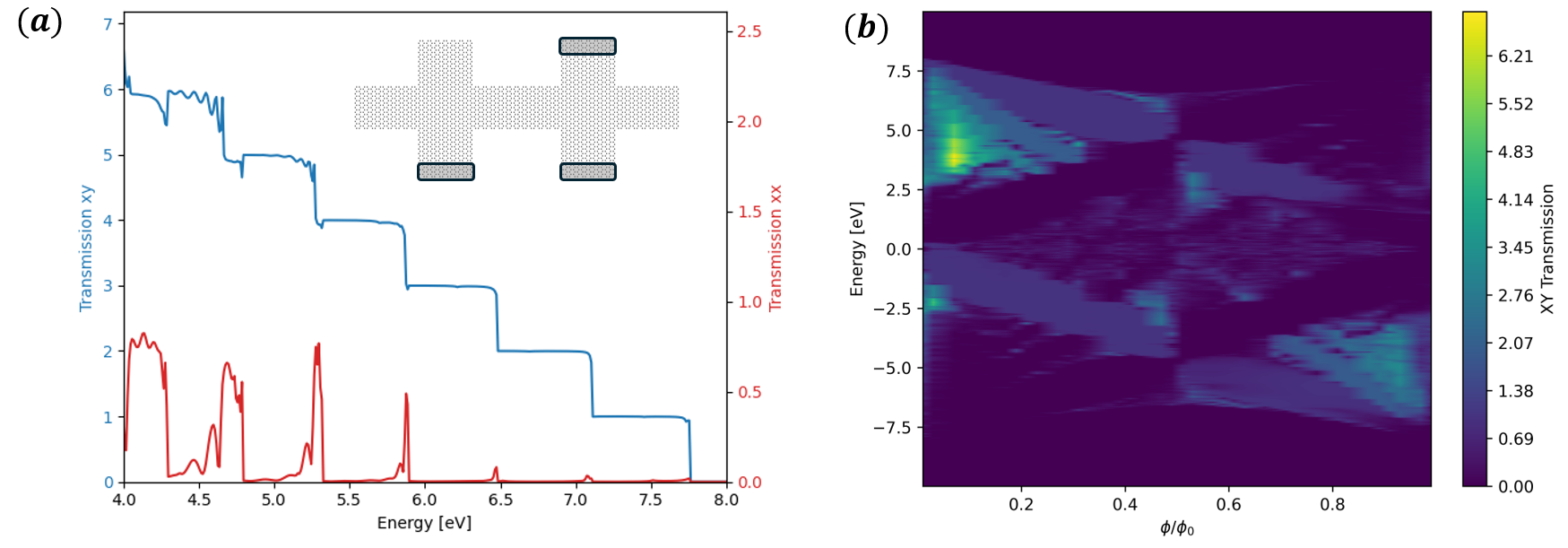}
    \caption{
    Integer quantum Hall effect in a Hall-bar geometry.
    (a) Longitudinal (red) and transverse (blue) transmission spectra
    for the graphene Hall-bar device shown in the inset.
    (b) Transverse transmission map as a function of magnetic flux and
    energy, recovering the Hofstadter-fractal signature.
    }
    \label{fig:hallbar_iqhe}
\end{figure*}

\begin{figure*}[!tp]
    \centering
    \includegraphics[width=.9\textwidth]{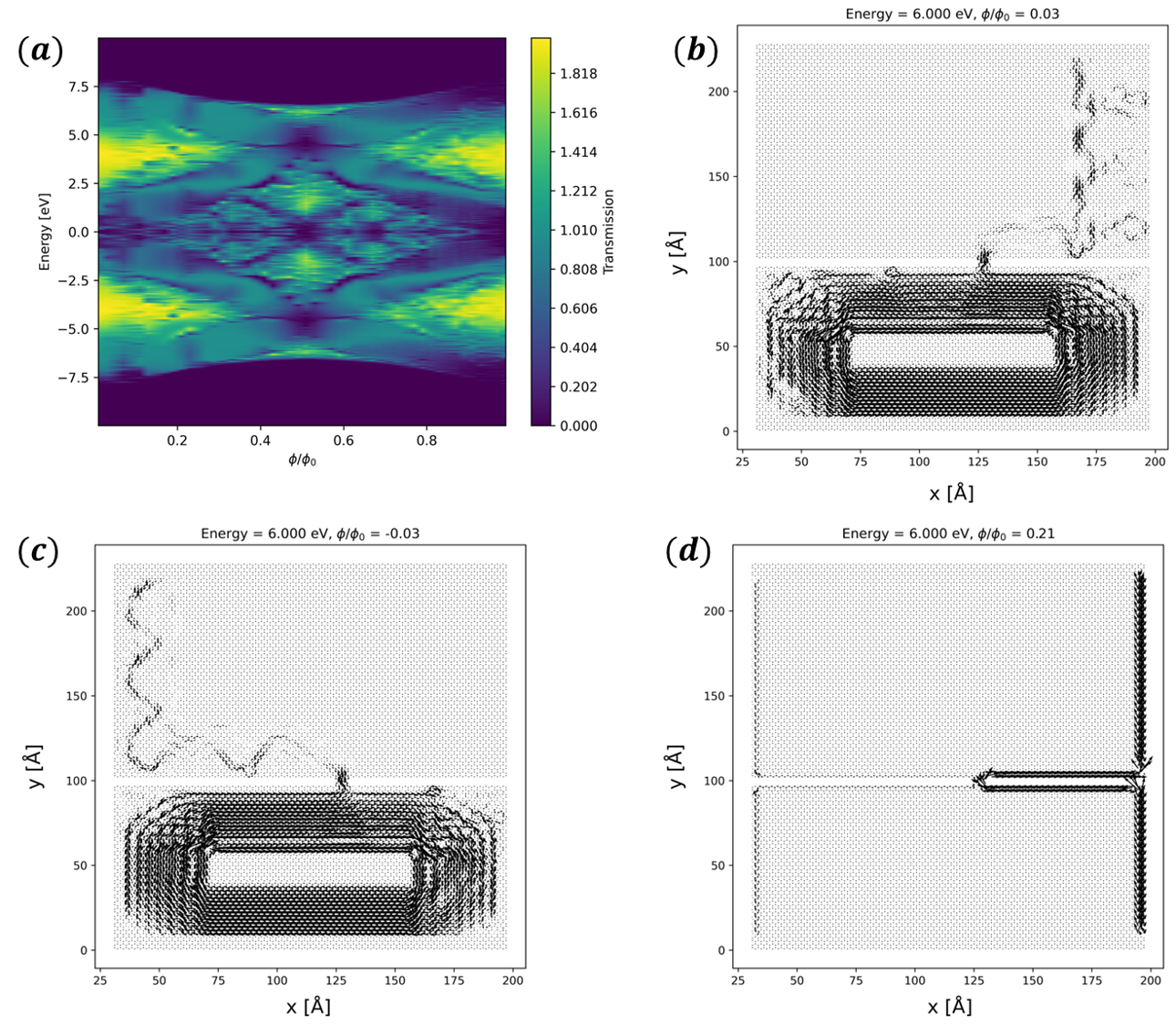}
    \caption{
    Magnetotransport in the slit-junction device.
    (a) Calculated magneto-transmission spectrum.
    (b),(c) Vector-transmission maps at $E=6\,\mathrm{eV}$
    illustrating chiral skipping-orbit behavior for opposite
    magnetic-field orientations.
    (d) Formation of edge-state-dominated transport at high magnetic flux.
    }
    \label{fig:slit_magnetotransport}
\end{figure*}

\begin{figure*}[!tp]
    \centering
    \includegraphics[width=.85\textwidth]{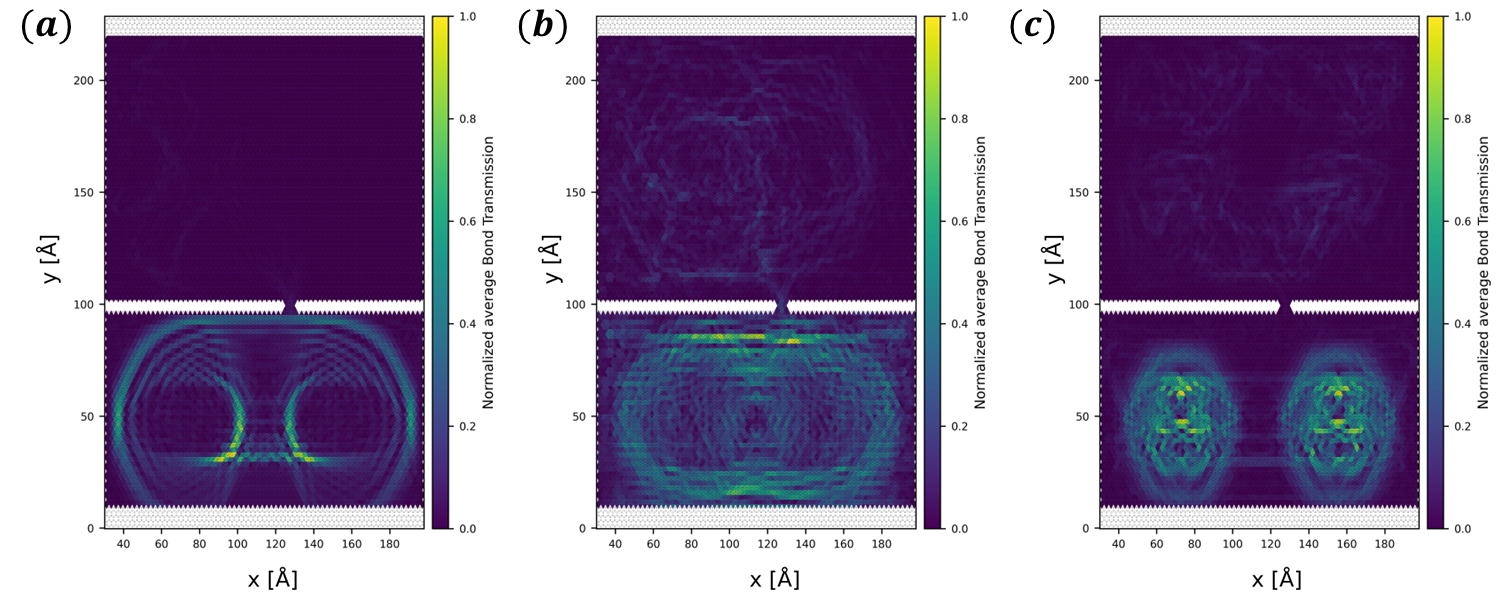}
    \caption{
    Quantum-scar-like states in the slit junction.
    Bond-transmission maps at selected energies, revealing stable
    recurring trajectories reminiscent of quantum scars. (a) $E=-3.505\,\mathrm{eV}$, (b) $-2.915\,\mathrm{eV}$, and (c) $-2.405\,\mathrm{eV}$.
    }
    \label{fig:quantum_scars_bond}
\end{figure*}

\begin{figure*}[!tp]
    \centering
    \includegraphics[width=.8\textwidth]{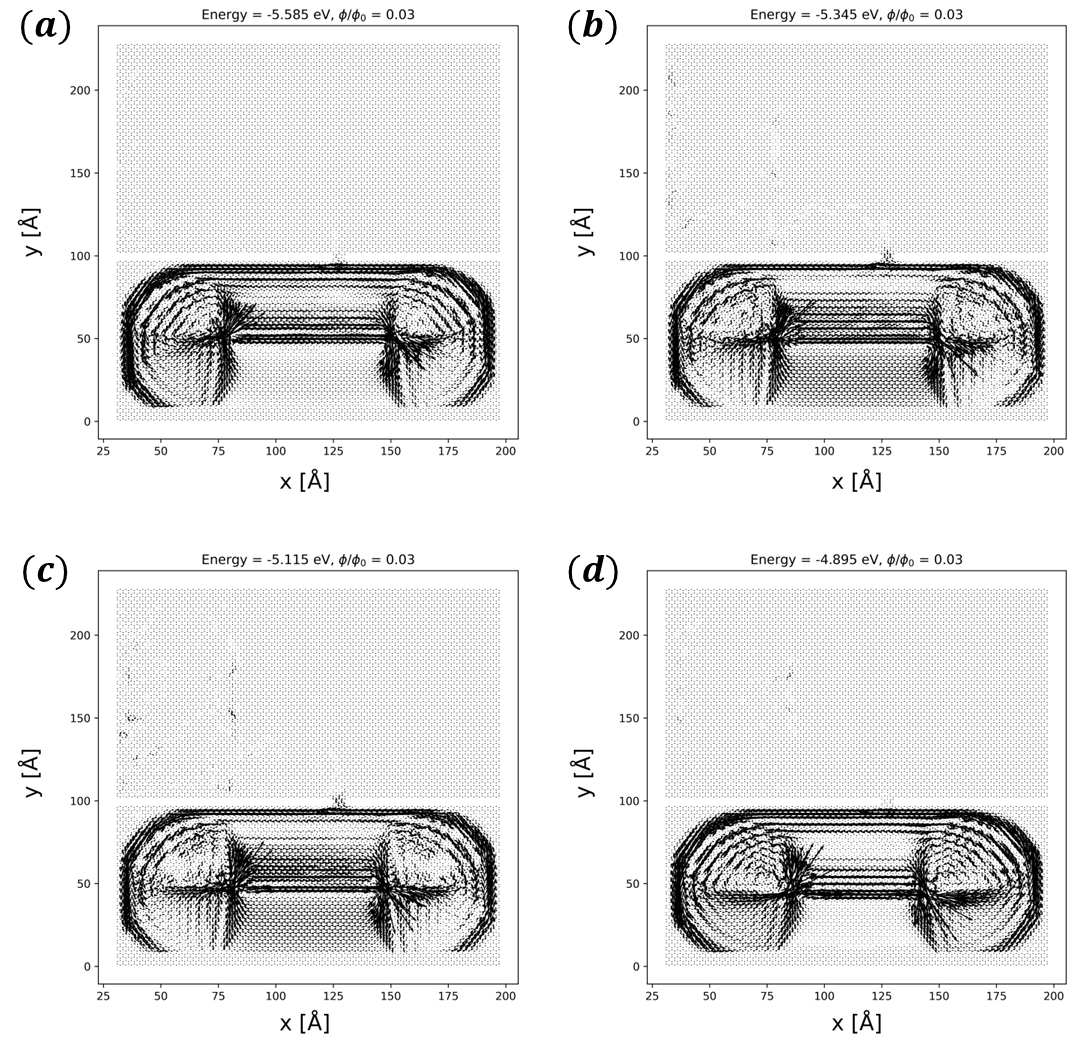}
    \caption{
    Additional scar-like transport patterns.
    Vector-transmission maps at $\Phi/\Phi_0=0.03$ and selected energies, illustrating the recurrence of stable transport trajectories.
    }
    \label{fig:quantum_scars_vector}
\end{figure*}

The magnetic field was included through the Peierls substitution
\cite{ref23}, whereby the vector potential is integrated along
each hopping path [Eqs.~(2)--(5) in the main text].

As an initial validation, the magnetic-field-induced modification of the band structure in graphene nanoribbons reported in Ref.~\onlinecite{Delplace2010} was reproduced. For this standalone band-structure benchmark, the Peierls phase is included directly in the periodic nanoribbon Hamiltonian. This differs from the transport calculations of the slit junction in the main text, where the magnetic field is restricted to the device region and the Peierls phases are not applied to the semi-infinite leads.
For the band-structure benchmark, the graphene unit-cell area
\begin{equation}
S=A_{\mathrm{UC}} = \frac{3\sqrt{3}}{2}a^{2},
\end{equation}
with $a$ the carbon--carbon distance, is taken as the relevant flux scale.
Figure~\ref{fig:magnetic_bands} shows the resulting dispersions for
armchair and zigzag nanoribbons, including the expected emergence of
Landau-level-like features.

The transport implementation was then verified by targeting the
Hofstadter butterfly in graphene \cite{Rhim2012}.
For this purpose, the graphene unit-cell area was kept as the magnetic
area entering the Peierls phase, so that Eq.~(5) of the main text reads
\begin{equation}
\phi_{ij} = \pi \frac{\Phi}{\Phi_{0}} \frac{(x_{i}-x_{j})(y_{i}+y_{j})}
{A_{\mathrm{UC}}},
\label{eq:peierls_uc}
\end{equation}
and the usual rational-flux condition was imposed,
\begin{equation}
\frac{\Phi}{\Phi_{0}} = \frac{p}{q},
\label{eq:rational_flux}
\end{equation}
where $p$ and $q$ are coprime integers.
The transmission map shown in Fig.~\ref{fig:hofstadter} reproduces the
expected Hofstadter spectrum \cite{Rhim2012}.

As an additional benchmark, a Hall-bar geometry was simulated and the integer quantum Hall effect was recovered. The longitudinal and transverse transmissions exhibit the expected
magnetic-field-induced structure, with a sequence of transmission peaks and quantized steps, respectively. The flux dependence of the transverse transmission again displays the Hofstadter signature [Fig.~\ref{fig:hallbar_iqhe}].

Having been validated, the toolbox was applied to the slit-junction device in the presence of a magnetic field.
Notably, complex absorbing potentials (CAPs) were not employed, in order to retain edge reflections. The device exhibits a rich magnetotransport spectrum consistent with previous studies \cite{Chakraborti2024}.
The spectrum is shown in Fig.~\ref{fig:slit_magnetotransport}(a). For low magnetic-field intensity, skipping-orbit behavior is observed [Fig.~\ref{fig:slit_magnetotransport}(b)], which reverses upon changing the sign of the magnetic field [Fig.~\ref{fig:slit_magnetotransport}(c)].
At sufficiently large magnetic flux, transport becomes dominated by robust edge states [Fig.~\ref{fig:slit_magnetotransport}(d)].

Additionally, recurring transmission patterns were observed at specific energies, as displayed in Figs.~\ref{fig:quantum_scars_bond} and \ref{fig:quantum_scars_vector}.
Their periodic-like trajectories resemble the quantum-scar-like states recently visualized in graphene quantum-dot $p$--$n$ junctions \cite{Ge2024} and analyzed theoretically in gate-defined bilayer-graphene cavities \cite{ref37}.

\clearpage

\section*{S2. Aharonov--Bohm Interferometer}

\begin{figure*}[!tp]
    \centering
    \includegraphics[width=.9\textwidth]{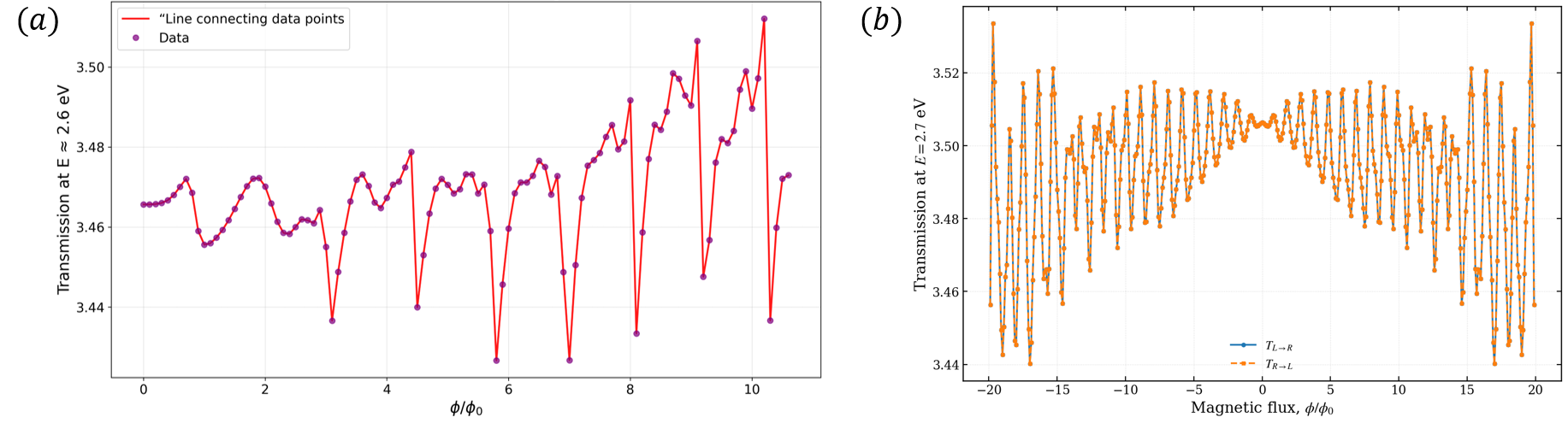}
    \caption{
    Additional validation of the Aharonov--Bohm response.
    (a) Transmission as a function of normalized magnetic flux
    $\Phi/\Phi_0$ at $E\simeq2.6\,\mathrm{eV}$.
    The oscillatory response is strongly reduced compared with that
    observed around $E=2.7\,\mathrm{eV}$, where the directional
    valley-polarized beams are more pronounced.
    (b) Numerical verification of Onsager--Casimir reciprocity at
    $E=2.7\,\mathrm{eV}$.
    The left-to-right transmission
    $T_{L\rightarrow R}(\Phi/\Phi_0)$ and the reciprocal
    right-to-left response
    $T_{R\rightarrow L}(-\Phi/\Phi_0)$ coincide over the calculated
    flux interval, in agreement with Eq.~\eqref{eq:onsager_flux}.
    }
    \label{fig:ab_area}
\end{figure*}

\begin{figure*}[!tp]
    \centering
    \includegraphics[width=.9\textwidth]{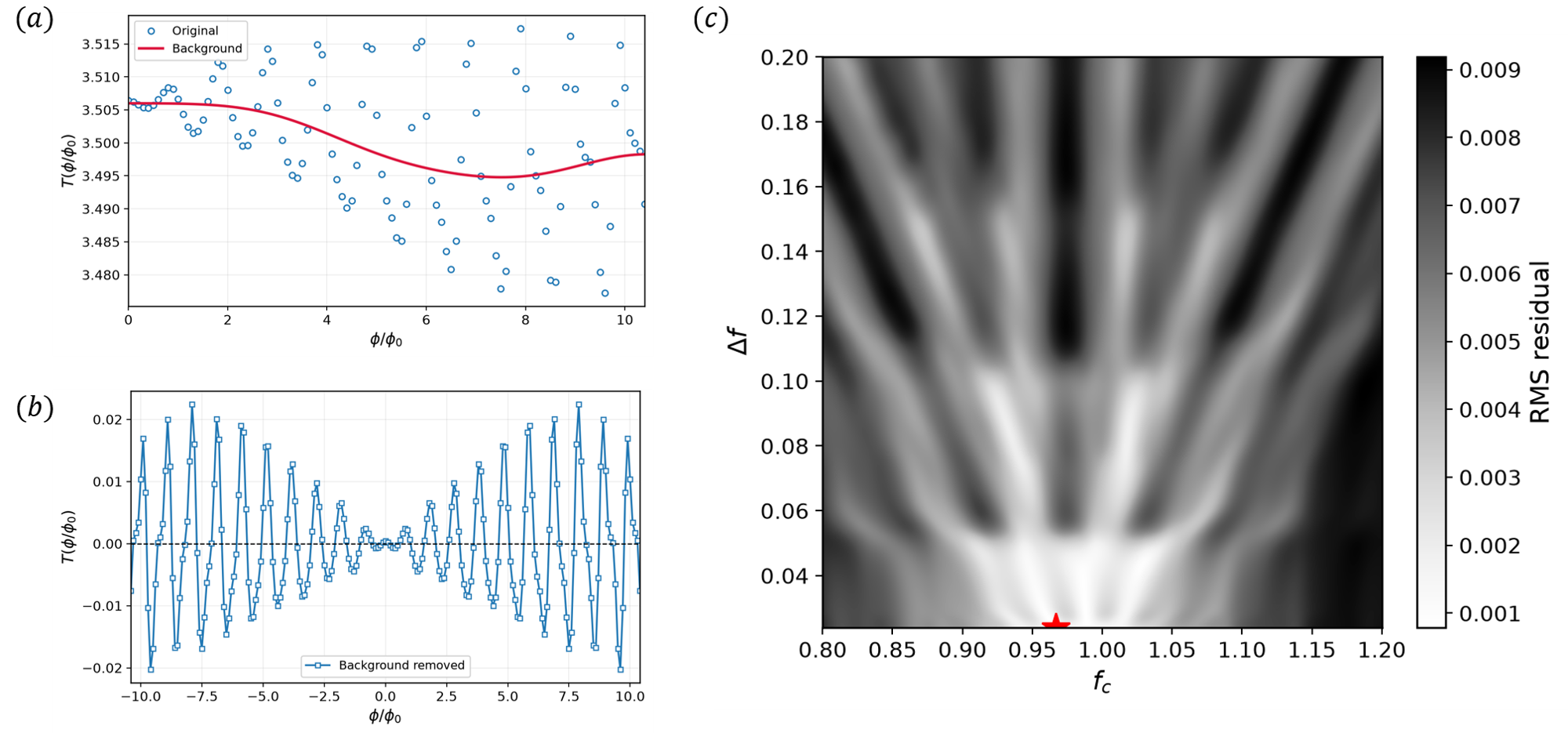}
    \caption{
    Fourier background removal and optimization of the
    split-frequency model for
    $W_{\mathrm{slit}}=10\,\text{\AA}$.
    (a): Original transmission and slowly varying background reconstructed from the removed Fourier components with
    $|f|<1/3$.
    (b): background-subtracted oscillatory transmission over the
    complete symmetric magnetic-flux interval.
    (c): RMS residual of the split-frequency model as a function of
    the central frequency $f_c$ and frequency separation $\Delta f$.
    At each point, the linear amplitude coefficients are determined by
    least-squares minimization.}
    \label{fig:fourier_filtering}
\end{figure*}

\begin{figure*}[!tp]
    \centering
    \includegraphics[width=.9\textwidth]{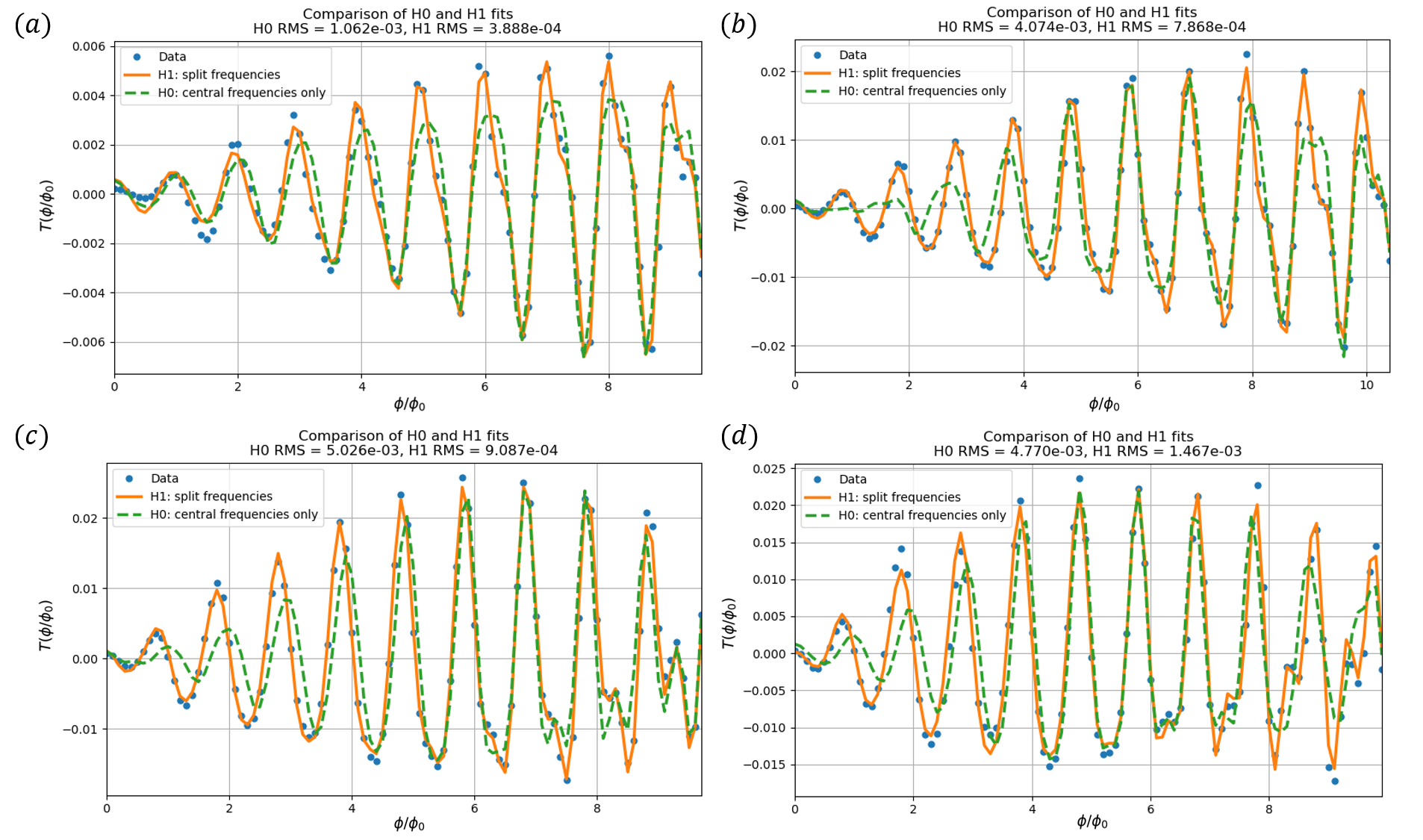}
    \caption{
    Comparison of the central-frequency (H0) and split-frequency (H1) models for different slit widths.
    Background-subtracted transmission for
    (a) $W_{\mathrm{slit}}=5\,\text{\AA}$,
    (b) $W_{\mathrm{slit}}=10\,\text{\AA}$,
    (c) $W_{\mathrm{slit}}=15\,\text{\AA}$, and
    (d) $W_{\mathrm{slit}}=20\,\text{\AA}$.
    Blue symbols denote the numerical transmission.
    Orange solid lines show the H1 model containing
    $f_m$, $f_c$, and $f_p$, together with their second harmonics.
    Green dashed lines show the H0 model containing only $f_c$ and
    $2f_c$, with higher-order even polynomial amplitude envelopes.
    The systematically smaller H1 residual demonstrates that the
    transmission modulation cannot be accounted for by amplitude
    variation alone and requires finite frequency splitting.
    }
    \label{fig:h0_h1_comparison}
\end{figure*}

\begin{figure*}[!tp]
    \centering
    \includegraphics[width=.9\textwidth]{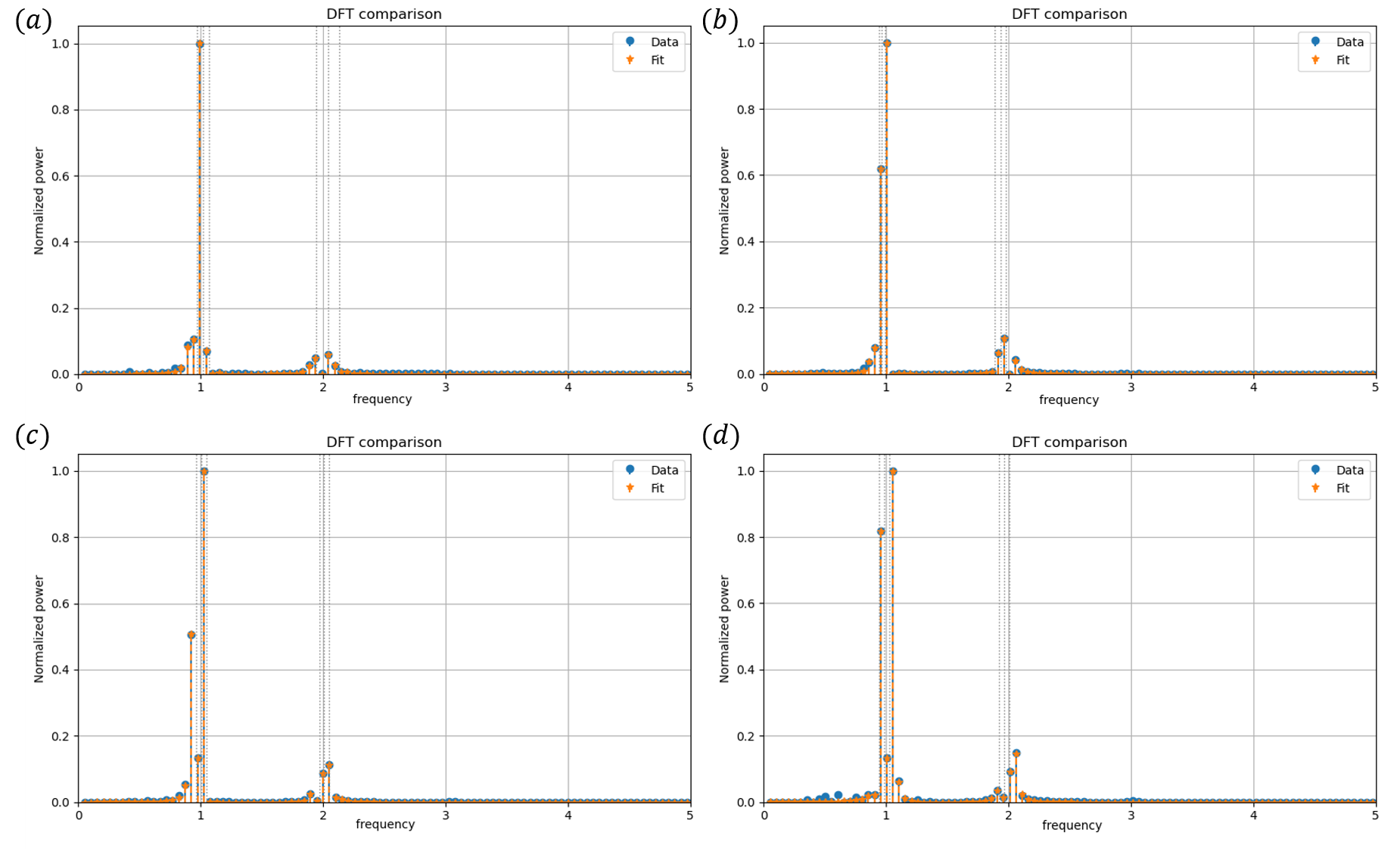}
    \caption{
    Frequency-domain validation of the split-frequency model for different slit widths.
    DFT power spectra of the background-subtracted numerical transmission
    (blue circles) and the corresponding best-fit H1 model (orange stars)
    for
    (a) $W_{\mathrm{slit}}=5\,\text{\AA}$,
    (b) $W_{\mathrm{slit}}=10\,\text{\AA}$,
    (c) $W_{\mathrm{slit}}=15\,\text{\AA}$, and
    (d) $W_{\mathrm{slit}}=20\,\text{\AA}$.
    Data and model are transformed using the same discrete normalized-flux
    grid and normalized to the maximum Fourier power of the numerical data.
    Gray dotted lines mark the fitted frequencies
    $f_m$, $f_c$, $f_p$, $2f_m$, $2f_c$, and $2f_p$.
    The agreement around both the fundamental and second-harmonic frequency
    groups confirms that the H1 model captures the dominant spectral
    content of the calculated transmission.
    }
    \label{fig:dft_model_comparison}
\end{figure*}

The interpretation of the slit junction as a two-path Aharonov--Bohm interferometer, together with the resulting flux- and gate-controlled oscillations, is presented in Sec.~III\,B of the main text.
Here, two additional validation tests of that response are provided, and the Fourier filtering and frequency-analysis procedures used in the main text are documented.

To confirm that the oscillatory response is closely related to the formation of strongly directional valley-polarized beams, the transmission was additionally calculated away from the energy at which the anisotropy is strongest.
Figure~\ref{fig:ab_area}(a) shows the transmission at
$E\simeq2.6\,\mathrm{eV}$.
Compared with the pronounced oscillations observed around $E=2.7\,\mathrm{eV}$ [Fig.~4(b) of the main text], the modulation is strongly suppressed, consistently with the weaker anisotropy of the electronic propagation at this energy.

A second consistency check is provided by the Onsager--Casimir reciprocity relation.
For a phase-coherent, nonmagnetic system in the linear-response regime, microscopic reversibility requires the transmission coefficients to obey
\begin{equation}
T_{L\rightarrow R}(B)
=
T_{R\rightarrow L}(-B),
\label{eq:onsager_casimir}
\end{equation}
or, in terms of the normalized magnetic flux,
\begin{equation}
T_{L\rightarrow R}
\left(
\frac{\Phi}{\Phi_0}
\right)
=
T_{R\rightarrow L}
\left(
-\frac{\Phi}{\Phi_0}
\right).
\label{eq:onsager_flux}
\end{equation}
Figure~\ref{fig:ab_area}(b) compares the left-to-right and right-to-left transmissions over the magnetic-flux interval
$-20\leq\Phi/\Phi_0\leq20$.
The two responses coincide after simultaneous reversal of the propagation direction and magnetic field, demonstrating that the numerical transport implementation satisfies the Onsager--Casimir reciprocity condition.
For the corresponding two-terminal conductance, this relation reduces to the familiar even-in-field symmetry $G(B)=G(-B)$ noted in the main text.

The transmission is evaluated at discrete and uniformly spaced values of the normalized magnetic flux.
A discrete Fourier transform (DFT) was therefore used to separate the slowly varying background from the oscillatory contribution.
The procedure is illustrated in Fig.~\ref{fig:fourier_filtering} for a slit width $W_{\mathrm{slit}}=10\,\text{\AA}$.

Prior to the Fourier transform, the mean transmission was subtracted in order to eliminate the residual zero-frequency component.
All Fourier coefficients satisfying
\begin{equation}
|f|<\frac{1}{3}
\label{eq:highpass_cutoff}
\end{equation}
were then set to zero. This high-pass filtering suppresses the slowly varying spectral contribution associated primarily with the magnetotransmission background, while preserving the frequency groups centered around the fundamental oscillation near $f\simeq1$ and its second harmonic near $f\simeq2$ [Fig.~6(a) of the main text].

An inverse DFT of the remaining coefficients yields the oscillatory component $T_{\mathrm{osc}}$.
The removed background is consequently obtained from
\begin{equation}
T_{\mathrm{bg}}
\left(
\frac{\Phi}{\Phi_0}
\right)
=
T
\left(
\frac{\Phi}{\Phi_0}
\right)
-
T_{\mathrm{osc}}
\left(
\frac{\Phi}{\Phi_0}
\right).
\label{eq:background_extraction}
\end{equation}
Figure~\ref{fig:fourier_filtering}(a) compares the
original transmission with the extracted slowly varying background.
Figure~\ref{fig:fourier_filtering}(b) shows the corresponding
background-subtracted oscillatory signal over the complete symmetric flux interval.

The frequencies entering the oscillatory model of Eq.~(22) of the main text were then determined by a two-dimensional scan over a central frequency $f_c$ and a frequency separation $\Delta f$.
For each pair $(f_c,\Delta f)$, the neighboring frequencies were defined
as
\begin{equation}
f_m=f_c-\Delta f,
\qquad
f_p=f_c+\Delta f,
\label{eq:sm_frequency_triplet}
\end{equation}
and the corresponding second harmonics $2f_m$, $2f_c$, and $2f_p$ were included as well, with
$f_i\in\{f_m,f_c,f_p,2f_m,2f_c,2f_p\}$.
For every tested pair $(f_c,\Delta f)$, the twelve linear coefficients of the amplitude envelopes [Eq.~(\ref{eq:H1_explicit})] were determined by least-squares minimization.
The quality of the reconstruction was quantified through the
root-mean-square residual
\begin{equation}
\mathrm{RMS}
=
\sqrt{
\frac{1}{N}
\sum_{n=1}^{N}
\left[
T_{\mathrm{osc},n}
-
T_{\mathrm{fit},n}
\right]^2
}.
\label{eq:sm_rms}
\end{equation}
The resulting RMS landscape is displayed in
Fig.~\ref{fig:fourier_filtering}(c).
The red star identifies the global minimum within the investigated parameter region.

To establish whether the observed modulation genuinely requires several nearby frequencies, rather than simply a strongly flux-dependent amplitude, two descriptions were compared.
Panel (b) of Fig.~\ref{fig:h0_h1_comparison} corresponds to the
$W_{\mathrm{slit}}=10\,\text{\AA}$ comparison shown in Fig.~6(d) of the main text; here the comparison is extended to all four slit widths, together with the corresponding residuals.

Throughout this subsection, the normalized magnetic flux is abbreviated as $q=\Phi/\Phi_0$.

The split-frequency model, denoted H1, contains the three fundamental frequencies $f_m=f_c-\Delta f$, $f_c$, and $f_p=f_c+\Delta f$ together with their second harmonics $2f_m$, $2f_c$, and $2f_p$.
Each component carries an even quadratic amplitude envelope, so that
\begin{align}
T_{\mathrm{H1}}(q)
={}&
\sum_{i\in\{m,c,p\}}
\left(c_{i,0}+c_{i,2}\,q^{2}\right)
\cos\!\left(2\pi f_i\, q\right)
\nonumber\\
&+
\sum_{i\in\{m,c,p\}}
\left(d_{i,0}+d_{i,2}\,q^{2}\right)
\cos\!\left(4\pi f_i\, q\right),
\label{eq:H1_full}
\end{align}
which coincides with Eq.~(22) of the main text.
Written out explicitly, H1 reads
\begin{align}
T_{\mathrm{H1}}(q)
={}&
\left(c_{m,0}+c_{m,2}q^{2}\right)\cos\!\left[2\pi (f_c-\Delta f) q\right]
\nonumber\\
&+\left(c_{c,0}+c_{c,2}q^{2}\right)\cos\!\left(2\pi f_c q\right)
\nonumber\\
&+\left(c_{p,0}+c_{p,2}q^{2}\right)\cos\!\left[2\pi (f_c+\Delta f) q\right]
\nonumber\\
&+\left(d_{m,0}+d_{m,2}q^{2}\right)\cos\!\left[4\pi (f_c-\Delta f) q\right]
\nonumber\\
&+\left(d_{c,0}+d_{c,2}q^{2}\right)\cos\!\left(4\pi f_c q\right)
\nonumber\\
&+\left(d_{p,0}+d_{p,2}q^{2}\right)\cos\!\left[4\pi (f_c+\Delta f) q\right].
\label{eq:H1_explicit}
\end{align}
It therefore contains twelve linear amplitude coefficients
$\{c_{i,0},c_{i,2},d_{i,0},d_{i,2}\}$ and the two nonlinear frequency parameters $f_c$ and $\Delta f$.

As an alternative hypothesis, H0 assumes that only a single fundamental frequency $f_c$ and its second harmonic $2f_c$ are present.
To provide H0 with substantial freedom to reproduce the varying
oscillation amplitude, its envelopes were expanded in even powers of $q$ up to twelfth order:
\begin{align}
T_{\mathrm{H0}}(q)
={}&
\left(\sum_{n=0}^{6} a_{2n}\, q^{2n}\right)
\cos\!\left(2\pi f_c\, q\right)
\nonumber\\
&+
\left(\sum_{n=0}^{6} b_{2n}\, q^{2n}\right)
\cos\!\left(4\pi f_c\, q\right),
\label{eq:H0_full}
\end{align}
Both models retain only cosine terms with even envelopes, so that $T(q)=T(-q)$, consistent with the symmetry of the calculated transmission under reversal of the magnetic flux.
For H0 the central frequency was re-optimized independently, by scanning $f_c$ over the interval $0.8\le f_c\le1.2$ and solving the linear least-squares problem for the envelope coefficients at each point.
Consequently, H0 contains the 14 linear coefficients
$\{a_{2n},b_{2n}\}$ and the single nonlinear parameter $f_c$, whereas H1 contains 12 linear coefficients and the two nonlinear parameters $f_c$ and $\Delta f$.
The comparison therefore does not favor H1 through a larger number of adjustable parameters.

Figure~\ref{fig:h0_h1_comparison} compares the two models for the four
slit widths, (a)~$W_{\mathrm{slit}}=5\,\text{\AA}$,
(b)~$10\,\text{\AA}$, (c)~$15\,\text{\AA}$, and (d)~$20\,\text{\AA}$.
The split-frequency H1 model yields a substantially smaller RMS
residual for every slit width, despite containing fewer free amplitude coefficients.
This indicates that the modulation cannot be reproduced solely by allowing the amplitude of a single-frequency oscillation to vary with magnetic flux.

The fitted model was additionally tested directly in Fourier space.
For both the background-subtracted numerical transmission and its best-fit H1 reconstruction, the residual mean was first removed.
A DFT was then calculated using exactly the same normalized magnetic-flux grid for the data and the model.

The Fourier powers were defined as
\begin{equation}
P(f)
=
|\widetilde{T}(f)|^{2},
\end{equation}
and both spectra were normalized by the maximum Fourier power of the numerical data.
This common normalization preserves the relative spectral weights and allows a direct comparison between the data and the fitted model.

Figure~\ref{fig:dft_model_comparison} shows the DFT comparison for the
same sequence of slit widths, (a)~$5\,\text{\AA}$, (b)~$10\,\text{\AA}$,
(c)~$15\,\text{\AA}$, and (d)~$20\,\text{\AA}$.
The fitted model reproduces the dominant group of Fourier components around $f\simeq1$, as well as the weaker group around the second harmonic $f\simeq2$.
The neighboring components within both frequency groups are also well reproduced.
The vertical dotted lines mark the continuously fitted frequencies $f_m$, $f_c$, $f_p$, $2f_m$, $2f_c$, and $2f_p$.

\section*{S3. Spin Polarization and Electrostatic Control}

\begin{figure*}[!tp]
    \centering
    \includegraphics[width=\textwidth]{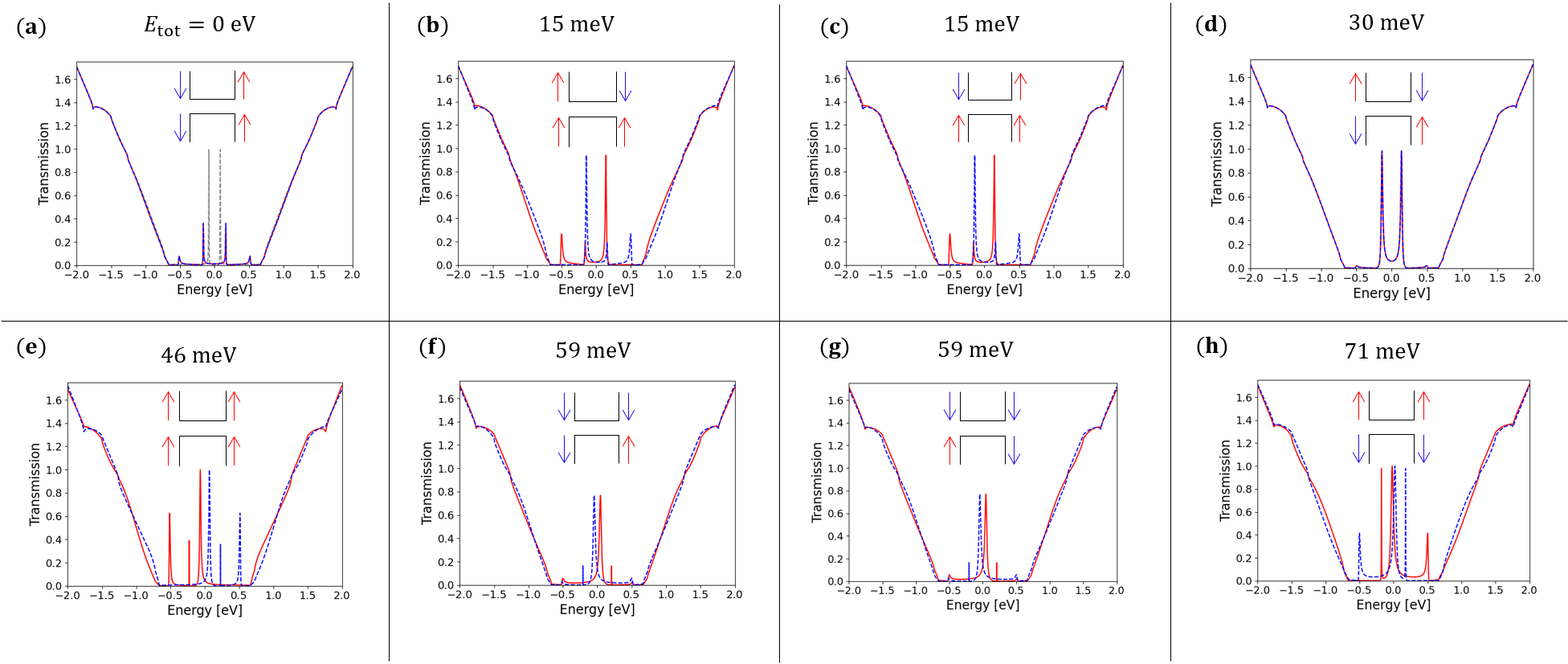}
    \caption{
    Self-consistent spin configurations of the slit junction within the mean-field Hubbard model.
    Each panel shows the spin-resolved transmission (solid red: spin up; dashed blue: spin down) for one converged configuration, with the corresponding edge-spin arrangement sketched in the inset and the total energy relative to the ground state given above the panel.
    The eight panels exhaust the eight arrangements that remain distinct after global spin inversion, ordered by increasing energy: (a) $0$ (antiferromagnetic ground state, taken as the reference), (b) $15$, (c) $15$, (d) $30$, (e) $46$ (ferromagnetic), (f) $59$, (g) $59$, and (h) $71\,\mathrm{meV}$.
    The pairs (b),(c) and (f),(g) are related by the left--right mirror operation $\mathcal{M}_{LR}$ and are therefore exactly degenerate.
    Spin-degenerate transmission occurs only for (a) and (d), the two configurations left invariant by
    $\mathcal{O}=\mathcal{M}_{LR}\otimes\mathcal{F}$, in agreement with Eqs.~(24) and (25) of the main text.
    }
    \label{fig:spin_configs}
\end{figure*}

\begin{figure*}[!tp]
    \centering
    \includegraphics[width=.9\textwidth]{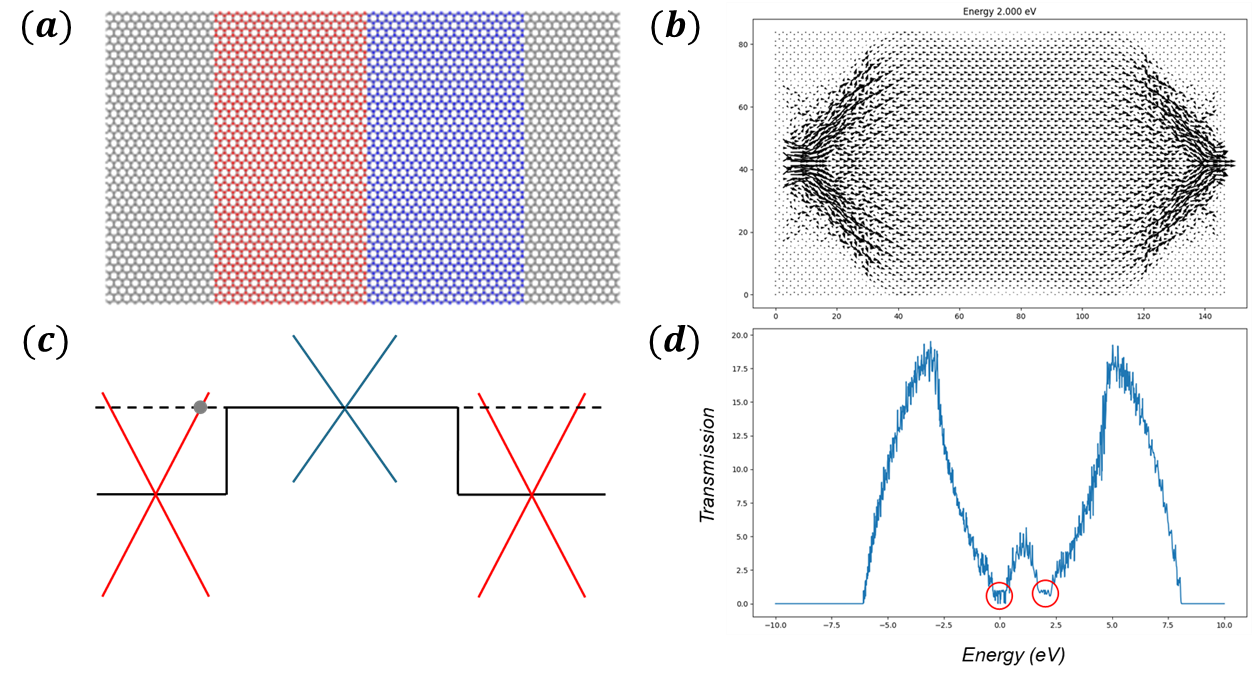}
    \caption{
    Electronic lensing in a graphene $p$--$n$ junction.
    (a) Device geometry with a symmetric electrostatic potential profile.
    (b) Vector-transmission map at $E=2\,\mathrm{eV}$ demonstrating Veselago focusing.
    (c) Schematic of the band alignment and effective refractive-index profile.
    (d) Transmission spectrum.
    }
    \label{fig:pn_lens}
\end{figure*}

\begin{figure*}[!tp]
    \centering
    \includegraphics[width=.9\textwidth]{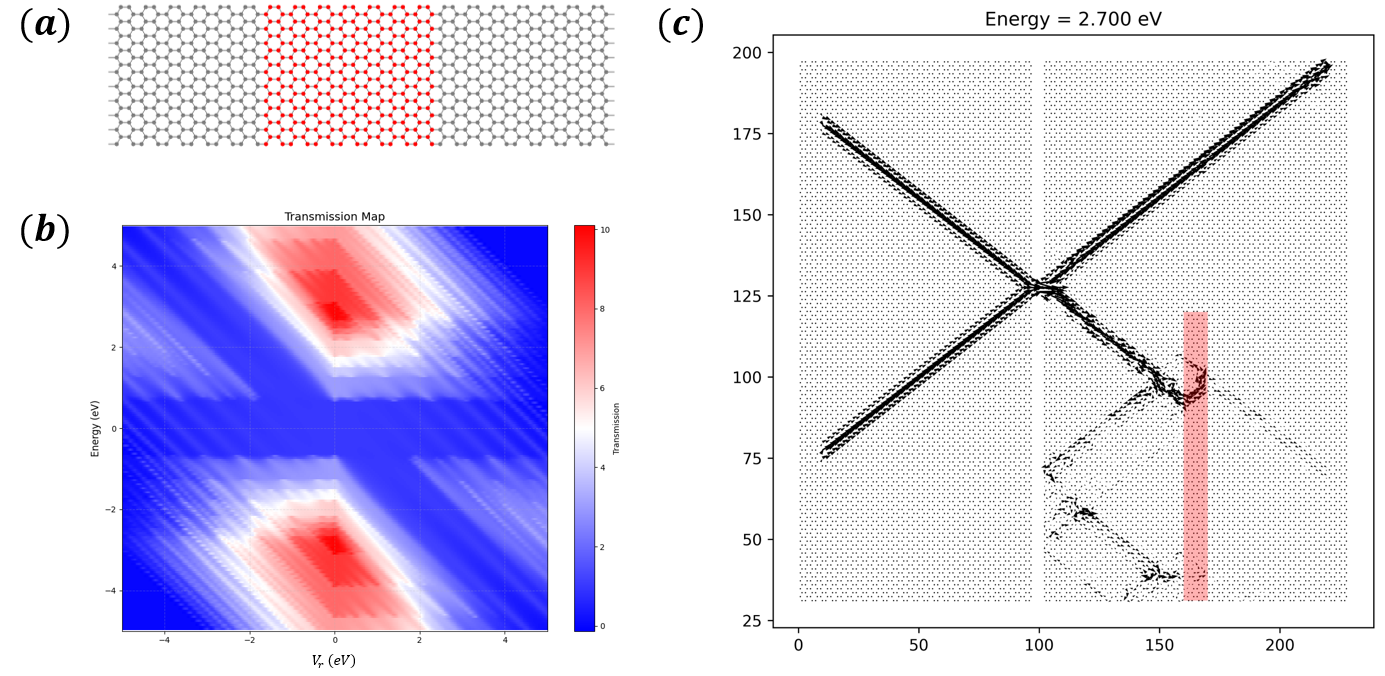}
    \caption{
    Electrostatic control of beam propagation.
    (a) Geometry with a central potential barrier (red region).
    (b) Transmission map as a function of carrier energy and barrier potential.
    (c) Vector-transmission map at $E=2.7\,\mathrm{eV}$ showing selective blocking of one valley-polarized beam by a local gate.
    }
    \label{fig:electrostatic_control}
\end{figure*}

Including Coulomb repulsion within the mean-field Hubbard model reveals spin polarization at the zigzag edges created by the slit \cite{ref14}.
The slit produces four such edges, so that the $2^4$ nominal
arrangements of the edge moments reduce, under global spin inversion, to eight distinct configurations.
All eight converge to self-consistent solutions and are collected in Fig.~\ref{fig:spin_configs}, ordered by increasing total energy, with the antiferromagnetic configuration identified as the ground state; the symmetry constraint governing their spin-resolved transmission is derived in Sec.~III\,C of the main text.

The role of symmetry is illustrated in Fig.~\ref{fig:spin_configs}.
Configurations (b) and (c), and likewise (f) and (g), are related by the left--right mirror operation $\mathcal{M}_{LR}$ alone; since $\mathcal{M}_{LR}$ is an exact symmetry of the device, each pair is degenerate and the two members share identical transmission spectra.
Only two of the eight configurations, (a) and (d), are invariant under the combined operation
$\mathcal{O}=\mathcal{M}_{LR}\otimes\mathcal{F}$ of Eqs.~(24) and (25) of the main text, and these are precisely the two whose spin-up and spin-down transmissions coincide; the remaining six all display spin-dependent transmission.
Equation~(24) of the main text is therefore verified across the complete set of self-consistent solutions rather than for isolated cases.

To induce spin filtering or phase delays, local electrostatic potentials defining $p$--$n$ junctions were introduced, as discussed in the main text [Eqs.~(8) and (9)].
A representative geometry is shown in Fig.~\ref{fig:pn_lens}(a), where onsite potentials are applied to the red and blue regions.

As a benchmark, Veselago focusing in graphene \cite{ref10} was first reproduced by choosing a symmetric potential profile $V_{\mathrm{red}}=V_{\mathrm{blue}}=2\,\mathrm{eV}$ on the two regions of Fig.~\ref{fig:pn_lens}(a).
The situation can be represented schematically in
Fig.~\ref{fig:pn_lens}(c) using the Dirac cones of graphene.
The two outer regions have their Dirac points at zero energy, while the central region has its Dirac point shifted to $E=2\,\mathrm{eV}$.
This modification is clearly observed in the transmission spectrum in Fig.~\ref{fig:pn_lens}(d), while the corresponding focusing effect is visible in the vector-transmission map in Fig.~\ref{fig:pn_lens}(b).

The transmission through a central electrostatic barrier,
shown in Fig.~\ref{fig:electrostatic_control}(a), was then mapped.
The resulting map as a function of carrier energy and barrier potential [Fig.~\ref{fig:electrostatic_control}(b)] displays the two bipolar regions of enhanced transmission expected when the barrier shifts the local Dirac point across the Fermi level, consistent with earlier literature \cite{Low2009}.
Finally, Fig.~\ref{fig:electrostatic_control}(c) shows a
vector-transmission map at $E=2.7\,\mathrm{eV}$ in which a local gate placed on the path of one of the two outgoing jets (red region) suppresses that arm while leaving the other intact, so that the gate acts as an electrostatic shutter for a single valley-polarized beam.

\section*{S4. Four-Band BLG Model and Valley-Resolved Current Maps}

\begin{figure*}[!tp]
    \centering
    \includegraphics[width=\textwidth]{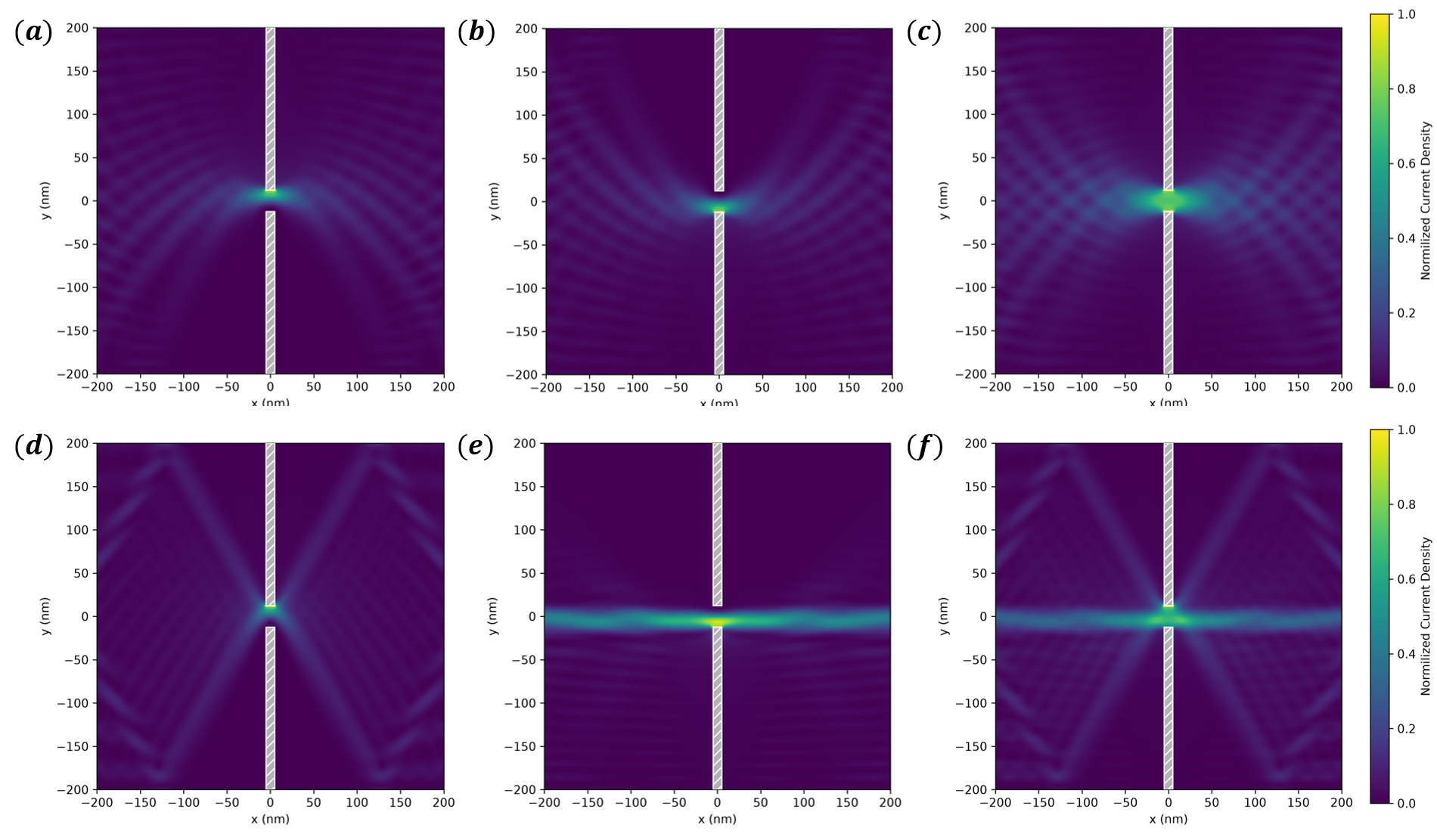}
    \caption{
    Normalized valley-resolved current maps of the BLG slit
    at $E=50\,\mathrm{meV}$ and $B=0$.
    Top row, (a--c): $\gamma_3=0$.
    Bottom row, (d--f): $\gamma_3=0.380\,\mathrm{eV}$ and a
    $90^\circ$ crystal-to-device orientation.
    From left to right, each row shows $K$, $K'$, and their vector-current sum.
    The nominal scattering region is $400\times400\,\mathrm{nm}^2$,
    with a $25\,\mathrm{nm}$ aperture in a $10\,\mathrm{nm}$-thick etched trench, shown by the hatched regions.
    The remaining model and device parameters are unchanged.
    Every map is normalized to its own maximum current magnitude; valley currents are added before taking the magnitude and normalizing.
    }
    \label{fig:blg_controls_sm}
\end{figure*}

To complement the low-energy BLG beam-formation results of Appendix~A in the main text, the four-band continuum model used by Schoeppl \emph{et al.}~\cite{ref37} was implemented in an open slit device. In the basis $(A_1,B_1,A_2,B_2)$, the Hamiltonian for valley $\xi=+1$ (K) or $\xi=-1$ (K$^\prime$) is
\begin{equation}
H_\xi=
\begin{pmatrix}
\epsilon_{A_1} & v_0\pi_\xi^\dagger & 0 & v_3\pi_\xi\\
v_0\pi_\xi & \epsilon_{B_1} & \gamma_1 & 0\\
0 & \gamma_1 & \epsilon_{A_2} & v_0\pi_\xi^\dagger\\
v_3\pi_\xi^\dagger & 0 & v_0\pi_\xi & \epsilon_{B_2}
\end{pmatrix},
\label{eq:sm_blg_hamiltonian}
\end{equation}
where $\epsilon_{A_1/B_2}=V\mp U_{\mathrm{BLG}}/2$, $\epsilon_{B_1/A_2}=V\mp U_{\mathrm{BLG}}/2+\Delta'$, $v_i=\sqrt{3}a_0\gamma_i/(2\hbar)$ for $i=0,3$, with lattice constant $a_0=0.246\,\mathrm{nm}$. The momentum is measured relative to the corresponding valley center, with $p_x=-i\hbar\partial_x$ and $p_y=-i\hbar\partial_y$. Rotation of the crystal relative to the fixed device axes is introduced through
\begin{equation}
\pi_\xi=(\xi\cos\alpha+i\sin\alpha)p_x
       +(-\xi\sin\alpha+i\cos\alpha)p_y.
\label{eq:sm_blg_rotation}
\end{equation}
Thus $\alpha=0^\circ$ gives $\pi_\xi=\xi p_x+ip_y$, whereas $\alpha=90^\circ$ gives $\pi_\xi=ip_x-\xi p_y$. 

The values $\gamma_0=3.16\,\mathrm{eV}$, $\gamma_1=0.381\,\mathrm{eV}$, $\Delta'=22\,\mathrm{meV}$, and $U_{\mathrm{BLG}}=40\,\mathrm{meV}$ are used. The warped case has $\gamma_3=0.380\,\mathrm{eV}$; the control sets only $\gamma_3$ to zero. The $\gamma_4$ term is omitted in both cases. Here $U_{\mathrm{BLG}}$ denotes the layer-asymmetry energy. The scalar potential is fixed at $V=0$ everywhere in the remaining device and both leads. All maps are evaluated at the stationary scattering energy $E=50\,\mathrm{meV}$.

The scattering region has nominal dimensions
$400\times400\,\mathrm{nm}^2$, with matching full-width semi-infinite leads attached on the left and right. A central etched trench of nominal thickness $10\,\mathrm{nm}$ leaves a $25\,\mathrm{nm}$ aperture.
The geometry and material parameters are identical for both
orientations.
The two physical valleys are treated independently; atomistic
intervalley scattering at the etched boundaries is not included.

The continuum Hamiltonian is implemented on an auxiliary square grid with spacing $b=1\,\mathrm{nm}$.
Following Ref.~\onlinecite{ref37}, a Wilson mass term with coefficient $w_D=47\,\mathrm{meV}$ is included. This numerical regularization shifts spurious high-momentum states away from the low-energy sector, suppressing contamination by the additional valleys generated by fermion doubling.

Figure~\ref{fig:blg_controls_sm}(a--c) shows the $\gamma_3=0$ control, with the same geometry and remaining material parameters. The separated two-beam pattern is no longer apparent in the valley sum. In Fig.~\ref{fig:blg_controls_sm}(d--f),
$\gamma_3=0.380\,\mathrm{eV}$ is restored and the crystal
orientation is changed to $\alpha=90^\circ$.
The valley-resolved maps reveal the contributions of $K$ and $K'$ to a three-beam pattern in the valley-summed current,
qualitatively similar to that obtained for monolayer graphene
in Fig.~13(b) of the main text.

\end{document}